# Flight masks of the Roman Space Telescope Coronagraph Instrument

A. J. Eldorado Riggs[a,*] Vanessa P. Bailey,[a] Dwight Moody,[a]
Kunjithapatham Balasubramanian,[a] Scott A. Basinger,[a] Ruslan Belikov,[b]
Eduardo Bendek,[a] John Debes,[c] Brandon D. Dube,[a] Jessica Gersh-Range,[d]
Tyler D. Groff,[e] N. Jeremy Kasdin,[d] Bertrand Mennesson,[a]
Brian Monacelli,[a] Douglas M. Moore,[a] Garreth Ruane,[a] Jagmit Sandhu,[a]
Fang Shi,[a] Erkin Sidick,[a] Nicholas Siegler,[a] Dan Sirbu,[b] John Trauger,[a]
Carey L. Weisberg,[a] Victor E. White,[a] Daniel W. Wilson,[a] Robert C. Wilson,[a]
Karl Y. Yee,[a] and Neil T. Zimmerman[e]

[a]Jet Propulsion Laboratory, California Institute of Technology, Pasadena, California, United States
[b]NASA Ames Research Center, Moffett Field, Mountain View, California, United States
[c]Space Telescope Science Institute, Baltimore, Maryland, United States
[d]Princeton University, Department of Mechanical and Aerospace Engineering, Princeton, New Jersey, United States
[e]NASA Goddard Space Flight Center, Greenbelt, Maryland, United States

**ABSTRACT.** Over the past two decades, thousands of confirmed exoplanets have been detected. The next major challenge is to characterize these other worlds and their stellar systems. Much information on the composition and formation of exoplanets and circumstellar debris disks can only be achieved via direct imaging. Direct imaging is challenging because of the small angular separations (<1 arcsec) and high star-to-planet flux ratios such as ∼$10^9$ for a Jupiter analog or ∼$10^{10}$ for an Earth analog in the visible. Atmospheric turbulence prohibits reaching such high flux ratios on the ground, so observations must be made above the Earth's atmosphere. The Nancy Grace Roman Space Telescope (Roman), planned to launch in late 2026, will be the first space-based observatory to demonstrate high-contrast imaging with active wavefront control using its Coronagraph Instrument. The instrument's main purpose is to mature the various technologies needed for a future flagship mission to image and characterize Earth-like exoplanets. These technologies include two high-actuator-count deformable mirrors, photon-counting detectors, two complementary wavefront sensing and control loops, and two different coronagraph types. We describe the complete set of flight masks in the Roman Coronagraph Instrument, their intended combinations, and how they were laid out, fabricated, and measured.

## 1 Introduction

Over 5700 exoplanets have been discovered to date, primarily via indirect methods that detect changes in the flux or Doppler shift of the host stars. The Decadal Survey on Astronomy and Astrophysics 2020 (Astro2020) laid out a long-term goal of discovering and characterizing habitable exoplanets around solar-type stars via direct imaging.[1] The direct imaging of exoplanets is challenging because of the high star-to-planet flux ratios at small angular separations. For example, the Sun and Earth viewed from 10 parsecs away have a planet-to-star flux ratio of $\approx 10^{-10}$ in the visible and a maximum separation of only 0.1 arcsec. Characterizing planets with such a high flux ratio in the visible will only be possible with a space-based observatory to avoid the severe performance limitations from atmospheric turbulence, dispersion, and absorption.

---

*Address all correspondence to A. J. Eldorado Riggs, aj.riggs@jpl.nasa.gov

In conjunction with a large space telescope for its collecting area and angular resolution, specialized optics are needed to suppress the bright glare of the starlight at the exoplanet's location. One of the leading technologies for this is the internal coronagraph. A coronagraph is a set of masks and/or mirrors that blocks or redirects the on-axis starlight and transmits off-axis sources such as exoplanets and debris disks. The 6-m Habitable Worlds Observatory (HWO) flagship proposed by Astro2020 is planned to use a coronagraph instrument to image and characterize Earth-like exoplanets around nearby stars.[1]

High-contrast coronagraphs with active wavefront control have been successfully demonstrated in laboratory testbeds but have not yet flown in space. To bridge the gap between current capabilities and HWO, the National Aeronautics and Space Administration (NASA) is including the Coronagraph Instrument as a technology demonstrator on the Nancy Grace Roman Space Telescope (Roman), a 2.4-m flagship observatory planned to launch no later than May 2027.[2–4] Some of the key technologies that the Roman Coronagraph Instrument will demonstrate are high-order wavefront sensing and control (HOWFSC) with two high-actuator-count deformable mirrors (DMs),[5–10] low-order wavefront sensing and control (LOWFSC) of pointing jitter and thermo-mechanical deformations of the observatory,[11–13] photon-counting detectors,[14] and two different types of coronagraphs.

The two complementary types of high-contrast coronagraphs included in the Roman Coronagraph Instrument are the Hybrid Lyot Coronagraph (HLC)[6,15–17] and the Shaped Pupil Coronagraph (SPC).[18–22] The HLC uses two DMs with large deformations, a phase-and-amplitude modifying focal-plane occulting spot, and a traditional Lyot stop to destructively interfere light and create a high-contrast region in the final image called the "dark hole." Much of the remaining starlight diffracts just outside the dark hole, so a field stop is needed with the HLC on Roman to limit the dynamic range on the detector and prevent ghosting through the color filter and imaging lens. The SPC uses a binary-amplitude apodizer to reshape the point spread function (PSF). Downstream, a focal plane mask (FPM) with a built-in field stop and then a Lyot stop block most of the rest of the starlight. Relatively little residual starlight remains at the final image plane, so dedicated field stops are not required for SPC imaging, unlike with the HLC. Because of the challenging telescope pupil obscurations (a large central obscuration and thick secondary mirror support struts) of the Roman Space Telescope, multiple coronagraphic mask configurations are needed to enable the desired high-contrast usage cases of (1) imaging close to the star, (2) performing spectroscopy close to the star, and (3) imaging over a wide field of view (FOV). Only use case 1 is officially required. The aspirational science cases enabled by these engineering goals are, respectively, to detect cool gas giant exoplanets and image inner debris disks, spectrally characterize the imaged exoplanets, and image the outer part of faint debris disks.

With this paper, we aim to provide a reference guide for the high-contrast imaging community. For potential users of the Roman Coronagraph Instrument, we provide a complete list of installed masks, their intended usage combinations, and their level of pre-flight testing and on-orbit support. To aid in the design of a future coronagraph instrument for HWO, we provide a narrative of key design decisions such as how the masks were chosen, how they were laid out, and what lessons we have learned so far. One major aspect of mask design that we do not include in this paper is the numerical optimization of the various high-contrast mask configurations. The general approaches used for those optimizations are already published in papers by Llop-Sayson et al.[23] and Gersh-Range et al.[24]

The structure of this paper is as follows. Section 2 provides a brief description of the instrument's optical layout. Section 3 gives a high-level explanation of how all the mask configurations were decided upon. To help the reader, there are diagrams and tables showing which masks are part of each mask configuration. A "mask configuration" is a specific set of masks in separate optical planes meant to be used together for starlight suppression or calibration. Next, mask layout considerations relevant to all the mask devices are described in Sec. 4. Sections 5–8 delve into the specifics of the mask layouts at each optical plane. Section 9 shows the final flight mask devices and key measurements of each. Finally, Sec. 10 details numerous lessons learned throughout the mask design and layout tasks. For reference, the color filters are listed in Appendix A. A list of all abbreviations used in the paper is provided in Appendix B for convenience.

## 2 Instrument Layout

Figure 1 labels the key optics and mechanisms within the instrument. The instrument receives the beam from the observatory near the middle of the optical bench at the fast steering mirror (FSM). After passing through the center of the second off-axis parabola (OAP) and off the focus control mirror (FCM), the beam gets modified by the two DMs. The beam next goes to the four optical planes for the coronagraphic masks. The first of those is the shaped pupil alignment mechanism (SPAM) which holds a fold mirror and the four reflecting-shaped pupil masks. Next is the focal plane alignment mechanism (FPAM) which carries all the diffractive FPMs. The beam path forks at the FPAM, with the rejected core of the stellar point spread function (PSF) sent through the low-order wavefront sensing (LOWFS) optical barrel element (LOBE) to the low-order wavefront sensing camera (LOCAM). The remainder of the starlight and any off-axis light is transmitted through the FPAM and continues on to and through the Lyot stop alignment mechanism (LSAM) and the field stop alignment mechanism (FSAM), the last of the two mask-carrying stages. From there, the light passes through the color filter alignment mechanism (CFAM) and then through either a lens, a spectroscopy prism, or a Wollaston prism at the dispersion polarization alignment mechanism (DPAM). The beam ends at the science detector, known as the exoplanetary systems camera (EXCAM). The six precision alignment mechanisms (PAMs) are

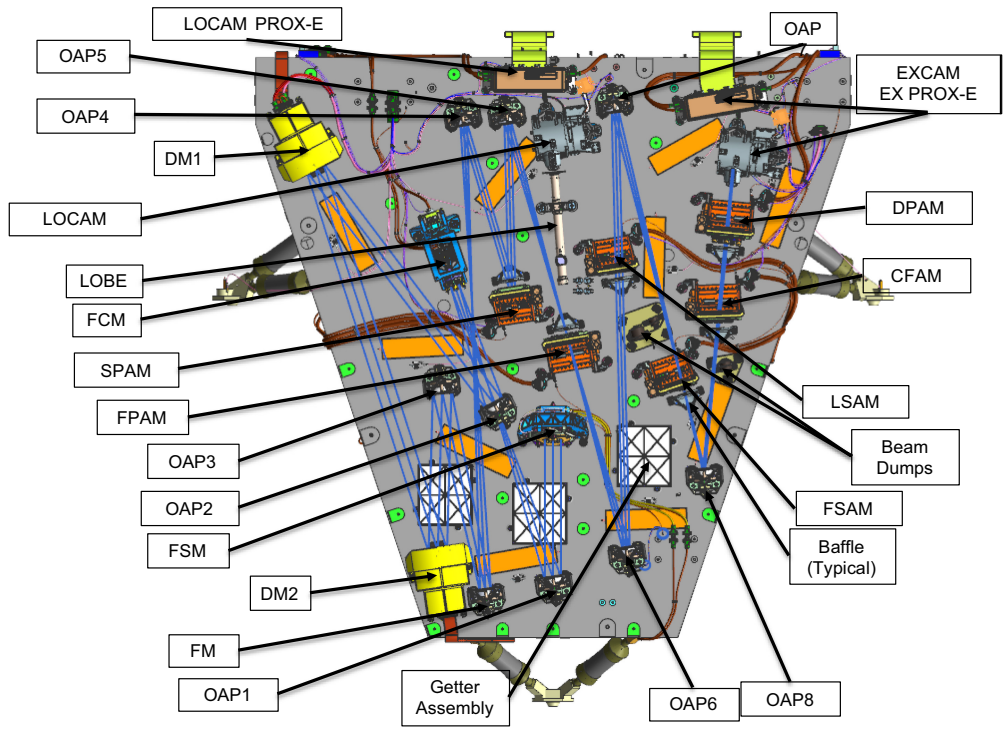

**Fig. 1** Optical layout of the Roman Coronagraph Instrument. The instrument receives light from the observatory at the fast steering mirror (FSM) and takes it all the way to the science detector or exoplanetary systems camera (EXCAM). All the coronagraphic masks are mounted on linear stages at the shaped pupil alignment mechanism (SPAM), focal plane alignment mechanism (FPAM), Lyot stop alignment mechanism (LSAM), and field stop alignment mechanism (FSAM). Other acronyms: LOCAM, low-order [wavefront sensing] camera; PROX-E, proximity electronics; OAP, off-axis parabola; DM, deformable mirror; LOBE, LOWFS Optical Barrel Element; FCM, focus control mirror; FM, fold mirror.

an international contribution from the Max Planck Institute for Astronomy (MPIA) in Heidelberg, Germany.[25]

The CFAM contains four main bandpass filters: band 1 is a 10.1% bandwidth centered at 575 nm, band 2 is a 17.0% bandwidth centered at 660 nm, band 3 is a 16.7% bandwidth at 730 nm, and band 4 is a 11.4% bandwidth at 825 nm. Bandwidth is defined here as the full width at half maximum (FWHM) of the filter transmission profile. Within each full band, there are also several narrower-bandwidth engineering filters used for HOWFSC or wavelength calibration. The complete list of CFAM filters and their designed bandpasses is in Table 5 in Appendix A.

The DPAM contains six standalone lenses, two polarization modules containing Wollaston prisms, and two spectroscopy modules containing Amici prisms.[26] One doublet lens is for focal plane imaging, another doublet is for pupil plane imaging, and the four singlet lenses provide defocus diversity for phase retrieval. The polarization modules are intended for use in bands 1 and 4 based on the supported coronagraphic mask configurations but could in principle also be used without performance degradation in bands 2 and 3. One spectroscopy module is designed for zero optical deviation and a spectral resolution of 50 at the center wavelength of band 2, and the other is designed to have those properties at the center of band 3. The spectroscopy modules could in principle be used in bands 1 and 4 as well but would have a small amount of optical deviation and different spectral resolutions.

## 3 Overview of All Mask Configurations

The Roman Coronagraph Instrument was originally planned as a technology demonstration instrument for high-contrast imaging, polarimetry, and low-resolution spectroscopy. In 2013, the Astrophysics-Focused Telescope Assets (AFTA) Coronagraph Working Group (ACWG) solicited proposals for different coronagraphic architectures and received six. The ACWG formulated and used a trade matrix to evaluate each coronagraphic architecture against required and desired scientific and engineering performance metrics. Three architectures were determined to have acceptable scientific yields and credible plans for reaching Technology Readiness Level (TRL) 5 by 2017: the HLC, SPC, and phase-induced amplitude apodization complex mask coronagraph (PIAACMC). The HLC and SPC were deemed a lower risk, while the PIAACMC offered a higher potential science yield. The HLC and SPC successfully reached TRL5 (i.e., better than $10^{-8}$ contrast in a 10% spectral bandwidth) and were therefore selected for flight.[27–29]

Both the HLC and SPC were kept because of their similar performance levels and complementary strengths and weaknesses, resulting in an instrument architecture similar to what is in the instrument now: an HLC for close-in imaging, an SPC for close-in spectroscopy but only over 1/3 the azimuthal field of view at a time, and another SPC for imaging out to the maximum DM-controllable separation.[29] Early Phase A designs included three orientations of the spectroscopy SPCs for full azimuthal coverage and versions of each coronagraph in several different bandpasses. Importantly, it was at this point that the PAMs that hold the various coronagraph masks were largely designed.

Over the next several years, for both technical and programmatic reasons, the required scope of the instrument was greatly simplified from several baseline and threshold technology requirements to just one, vestigially named Technology Threshold Requirement 5 (TTR5). The exact wording for TTR5 is: The Roman coronagraph shall be able to measure, with SNR ≥ 5, the brightness of an astrophysical point source located between 6 and 9 $\lambda/D$ from an adjacent star with a $V_{AB}$ magnitude ≤ 5, with a flux ratio ≥ $1 \times 10^{-7}$. The bandpass shall have a central wavelength ≤600 nm and a bandwidth ≥10%. This threshold requirement was intentionally set comfortably above the current best estimates of instrument performance on orbit of slightly better than $10^{-8}$ contrast. To meet this single remaining top-level requirement, only one coronagraphic mask configuration was technically necessary, and the HLC band 1 imaging was the only one to meet TTR5 at that time. The instrument's optical design was already largely mature, so the project was directed to maintain hardware for most of the originally planned capabilities on a "best-effort" basis while staying within the established budget and schedule. In addition, although several of the coronagraph mask types originally envisioned in the early Phase A design had been descoped, their unused mounting slots in the PAMs were maintained. As we will discuss in the following section, this enabled a later contribution of mask hardware by NASA's Exoplanet Exploration Program (ExEP).

### 3.1 Classes of Mask Configurations

There are three different classes of coronagraphic mask configurations in the as-built Roman Coronagraph Instrument: required, best effort, and unsupported. These configurations differ based on their levels of testing and software support.

Figure 2 provides a visual guide to the key optical elements in each high-contrast mask configuration. A simplified version of the optical train is shown at the top to make clear which elements are in the collimated beam and which are at a focal plane. All the pupil masks and DM surfaces are shown to scale relative to each other in terms of beam diameter. All the FPMs, field stops, and PSFs are shown to scale relative to each other in terms of diffraction widths (i.e., the wavelength divided by the telescope diameter, $\lambda/D$). The PSFs displayed include the field stop for the HLCs but do not for the SPCs because the slits or multi-star square field stops could be placed in many possible locations. Krist et al.[30] provide detailed analysis and plots of the stellar dark holes and off-axis PSFs for required and contributed mask configurations, excluding band 2 spectroscopy, which only later became supported. There was no analogous modeling for the unsupported mask configurations except for multi-star imaging, which was performed outside of the project.[31,32]

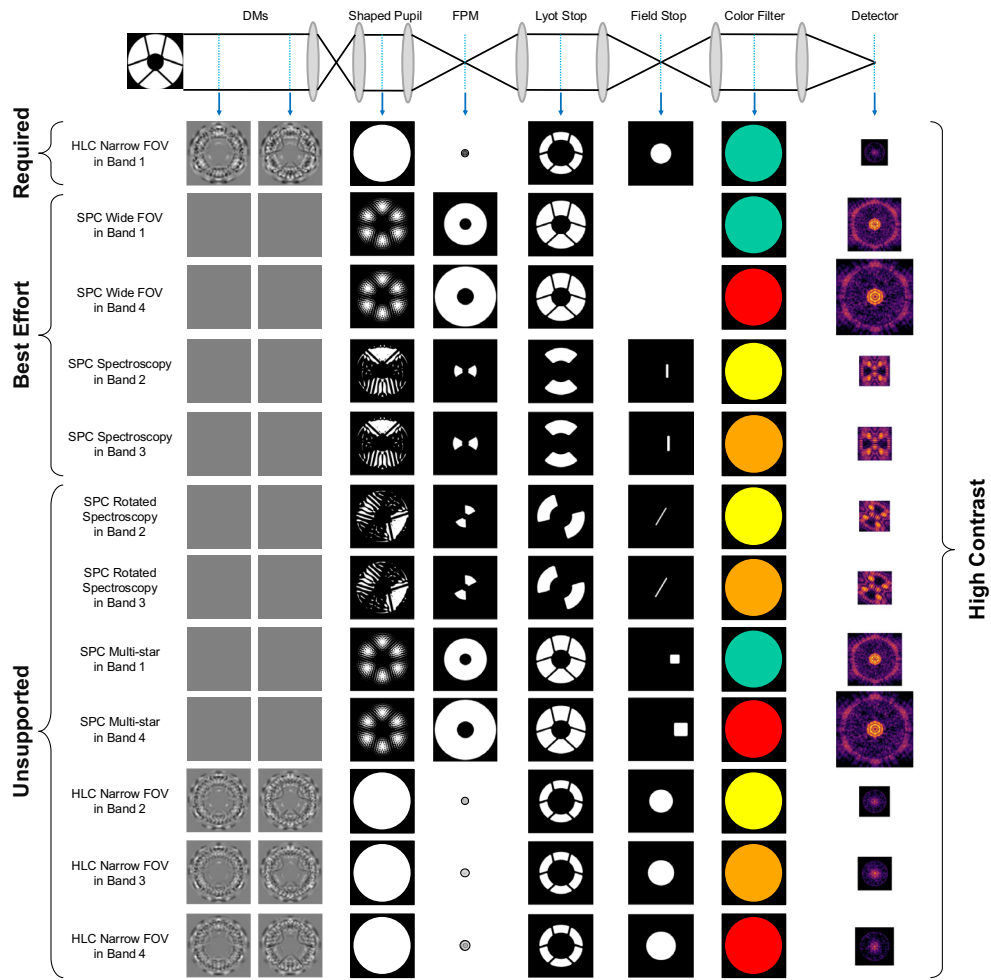

**Fig. 2** Masks and filters in all the designed high-contrast mask configurations of the Roman Coronagraph Instrument. In general, many field stops could be used with each mask config, but only the nominal field stop for each config is shown above. The DM surfaces shown exclude the contribution needed to correct aberrations in the optical system. The multi-star imaging dark holes are shown for a single star only, as any two-star geometry would be case-dependent. All the focal plane masks, field stops, and PSFs are shown to scale relative to each other in terms of $\lambda/D$. All pupil masks and DM surfaces are shown to scale relative to each other in terms of beam diameter.

The one required mask configuration, the HLC in band 1, fulfills TTR5 and therefore is fully supported and received end-to-end, system-level testing during the instrument's pre-delivery thermal vacuum (TVAC) performance testing at NASA's Jet Propulsion Laboratory (JPL) in spring 2024.[33,34] Note that only imaging is required for TTR5, so polarimetry in band 1 is only supported on a best-effort basis. This mode is guaranteed to be supported on-orbit.

Next, the four best-effort mask configurations are all SPCs. These serve to perform spectroscopy in bands 2 and 3 as well as wide FOV imaging and polarimetry in bands 1 and 4. They received component-level hardware support and full instrument software support, although observational software support is on a best-effort basis. The band 1 wide FOV SPC is also capable of accomplishing TTR5, so it was the only other mask configuration tested at the system level in TVAC.[33] These modes are not guaranteed time on-orbit, although their level of maturity makes them more likely to be exercised on-orbit than the "unsupported" modes we discuss next.

NASA's ExEP contributed hardware for more than a dozen unsupported mask configurations as well as the hardware for the now-best-effort band 1 wide FOV SPC. They were selected in the spring of 2020 as the best additional mask configurations by a working group of scientists and engineers at institutions across the USA. These unsupported masks are situated in otherwise unused space alongside the supported masks. Other than being installed and aligned, they have not received any hardware testing or software support. If resources allow, one or more of these modes may be commissioned in the 90-day baseline observation phase; this decision will be made as part of future observation planning.

The unsupported mask configurations are divided into three sub-categories: high-contrast, low-contrast, and calibration. The unsupported high-contrast configurations are mostly chosen to fill gaps in the covered parameter space of capabilities; for example, providing narrow FOV imaging to bands 2, 3, and 4 and one other orientation of the spectroscopy SPC in bands 2 and 3. Because the general alignment, calibration, and HOWFSC steps for these high-contrast unsupported mask configurations are essentially the same as those for the required and contributed ones, there are no changes needed to flight software and minor changes needed to the ground software needed to use them. One major exception is for the multi-star wavefront control (MSWC) mask configurations, for which substantial testbed, ground software, and flight software development would be necessary.

The unsupported band 2 and 3 HLCs could also be used for spectroscopy but are considered backups. They have worse design contrast (i.e., close to $10^{-8}$ versus $\approx 10^{-9}$ for the spectroscopy SPCs) because the inherently chromatic DM-generated contrast of the HLC is worse in broader bandwidths. In addition, they are more sensitive to low-order aberrations, which is acceptable over the few hours needed for imaging but not for the hundreds of hours needed for spectroscopy. A secondary concern is that the HLCs for the broader bandpasses and longer wavelengths of bands 2 and 3 require significantly more DM stroke—nearly maxing out the DM allocation—whereas the SPCs need no DM stroke in their nominal designs.

The unsupported low-contrast configurations (shown in Fig. 3) are all conventional Lyot coronagraphs with large diameter occulters that can utilize the instrument's full field of view, so they could be used equally well in multiple bandpasses, limited mainly by the anti-reflective (AR) coating of the FPAM substrate. One should note that smaller HLC occulters—when used with DMs giving a flat phase—can also be used as conventional Lyot coronagraphs. The low-contrast Lyot coronagraphs can be used with either Lyot stop that has struts blocking those in the telescope pupil, but most users would probably prefer the one for the wide FOV SPC because it provides higher throughput and a sharper PSF at the expense of slightly worse contrast. The low-contrast mask configurations can fully re-use the ground and flight software already developed for the required mask configuration.

Finally, the unsupported calibration masks (Fig. 3) are dual-path Zernike wavefront sensors (ZWFS) to enable sensitive pupil-plane wavefront measurements as desired for HWO.[35] Every FPM is already used as a low-order ZWFS in reflection, but these contributed ZWFS dielectric dimples are unique in that they were designed to work as a ZWFS concurrently along both the reflected LOCAM beam path to maintain pointing control and along the transmitted EXCAM beam path to provide high spatial frequency phase measurements. The existing LOWFSC flight software can lock the line-of-sight (LOS) loop if the mask starts close enough, but the LOS capture range and coarse alignment error after a move from a known FPAM position have yet to be studied. A high-level concept of operations was proposed as part of the contributed mask working group process, but the detailed operational implementation, data processing strategies, and ground software for the dual-path ZWFS would require a substantial effort.

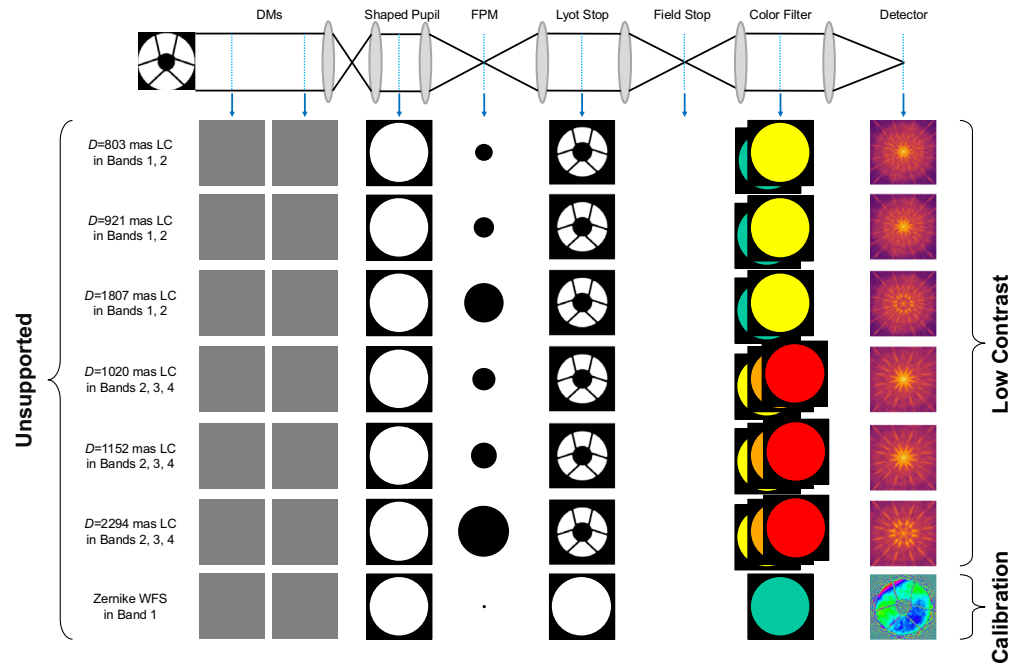

**Fig. 3** Masks and filters in all the unsupported low-contrast and calibration mask configurations of the Roman Coronagraph Instrument. Any or no field stop could be used. All the focal plane masks and PSFs are shown to scale relative to each other (in terms of $\lambda/D$). All pupil masks and DM surfaces are shown to scale relative to each other (in terms of beam diameter).

### 3.2 Designed Mask Configurations

To use the instrument, it is useful to have the position names for the masks and filter in each of the designed mask configurations. These are provided in Table 1 for all the high-contrast mask configurations and in Table 2 for the rest of the unsupported ones. The columns show the classifications, key optical elements, and FOV parameters for each mask configuration. Each row in Tables 1 and 2 is for a different mask configuration. The PAM position names in these tables refer to the figures shown in later sections with labeled mask layouts for each PAM. The small number of pupil plane positions at SPAM and LSAM are named after the configuration type or "OPEN" for the clear aperture positions. The FPAM and FSAM mask arrays are much more densely packed, so their mask names refer to locations in grids with "R" denoting the row number and "C" the column number. FPAM has masks on four substrates, so those position names also have prefixes of "HLC12," "HLC34," "SPC12," or "SPC34" denoting whether the substrate is for HLC or SPC only and whether it is for bands 1 and 2 or for bands 3 and 4.

The rows of Table 1 match up to the rows in Fig. 2 except for the "alternate" HLC FPMs in bands 2 and 4. These alternates are the original FPM designs with complicated occulter profiles for high contrast. However, it was discovered after the FPM substrate HLC12 was finished that the capture ranges for these two FPMs were much smaller (45 mas) than desired (80 mas) for the LOWFSC pointing control loop. Luckily, the anti-reflection (AR) coating on the HLC34 FPM substrate also transmits band 2 light; this substrate had not yet been fabricated and had space remaining for several more FPMs. New band 2 and 4 HLC occulters were designed with simpler, inverted-top-hat dielectric profiles with simulated LOWFSC performance meeting the pointing control capture range needs. These FPMs were fabricated on the HLC34 substrate along with the original HLC band 3 and 4 FPMs. Several original (now alternate) design band 4 HLC FPMs were also fabricated on the substrate in case the on-orbit pointing control capture range turns out to be sufficiently low.

**Table 1** Table of all the designed high-contrast mask configurations in the Roman Coronagraph Instrument. Mask PAM positions reference the labeled mask array layouts shown in later sections. As an example for FPAM, HLC12_C2R1:9 means column 2 and rows 1 to 9 on FPAM substrate HLC12. For the SPC multi-star, the FOV is set by the field stop opening but can be placed anywhere within the listed annular FOV. Abbreviations: Azim., azimuthal (see text for explanation); IWA, inner working angle; OWA, outer working angle.

| Mask configuration class | Coro. type | Configuration name | Band number | SPAM position | FPAM position | LSAM position | FSAM position | Azim. FOV (deg) | IWA ($\lambda/D$) | IWA (mas) | OWA ($\lambda/D$) | OWA (mas) |
|---|---|---|---|---|---|---|---|---|---|---|---|---|
| Required | HLC | Narrow FOV | 1 | OPEN | HLC12_C2R1:9, HLC12_C3R1:8, HLC12_C4R7:9 | NFOV | R1C1, (R1C3), (R6C6), R5C3, R5C4 | 360 | 3.0 | 150.6 | 9.7 | 486.8 |
| Best effort | SPC | Wide FOV | 1 | WFOV | SPC12_R1C1:2, SPC12_R2C1 | WFOV | open, R1C5, R6C1, R6C2 | 360 | 5.9 | 296.1 | 20.1 | 1008.8 |
|  | SPC | Wide FOV | 4 | WFOV | SPC34_R5C1, SPC34_R8C1, SPC34_R8C3 | WFOV | open, R1C5, R6C1, R6C2 | 360 | 5.9 | 424.9 | 20.1 | 1447.4 |
|  | SPC | Spectroscopy | 2 | SPEC | SPC12_R3C2, SPC12_-R4C2, SPC12_-R4C4 | SPEC | open, R2C5, R3C1, R3C2, R6C5 | 130 | 3.0 | 172.8 | 9.1 | 524.2 |
|  | SPC | Spectroscopy | 3 | SPEC | SPC34_R2C2, SPC34_R4C3, SPC34_R6C1 | SPEC | open, R1C2, R3C1, R3C2, R4C2, R4C6 | 130 | 3.0 | 191.2 | 9.1 | 579.8 |
| Unsupported (high contrast) | SPC | Rotated spectroscopy | 2 | SPECROT | SPC12_R3C1, SPC12_R3C3, SPC12_R4C3 | SPECROT | R4C3, R4C4 | 130 | 3.0 | 172.8 | 9.1 | 524.2 |
|  | SPC | Rotated spectroscopy | 3 | SPECROT | SPC34_R2C3:4, SPC34_R4C2 | SPECROT | R2C2, R4C2 | 130 | 3.0 | 191.2 | 9.1 | 579.8 |
|  | SPC | Multi-star | 1 | MSWC | SPC12_R1C1:2, SPC12_R2C1 | WFOV | R4C1 | 360 | 5.9 | 296.1 | 20.1 | 1008.8 |
|  | SPC | Multi-star | 4 | MSWC | SPC34_R5C1, SPC34_R8C1, SPC34_R8C3 | WFOV | R1C4, (R4C1) | 360 | 5.9 | 424.9 | 20.1 | 1447.4 |
|  | HLC | Narrow FOV | 2 | OPEN | HLC34_R7C1:2, HLC34_R8C2:3 | NFOV | R3C3, R2C3 | 360 | 3.0 | 172.8 | 9.7 | 558.8 |
|  | HLC | Narrow FOV, alternate | 2 | OPEN | HLC12_C4R1:6, HLC12_C5R1:8 | NFOV | R3C3, R2C3 | 360 | 3.0 | 172.8 | 9.7 | 558.8 |
|  | HLC | Narrow FOV | 3 | OPEN | HLC34_R5C1:3, HLC34_R6C1:3 | NFOV | R3C4, R4C5, R5C1, R5C2 | 360 | 3.0 | 191.2 | 9.7 | 618.1 |
|  | HLC | Narrow FOV | 4 | OPEN | HLC34_R3C1:3, HLC34_R4C1:3 | NFOV | R3C5 | 360 | 3.0 | 216.0 | 9.1 | 655.3 |
|  | HLC | Narrow FOV, alternate | 4 | OPEN | HLC34_R2C1:3 | NFOV | R3C5 | 360 | 3.0 | 216.0 | 9.1 | 655.3 |

**Table 2** List of designed mask configurations for low-contrast or calibration measurements in the Roman Coronagraph Instrument. Mask PAM positions reference the labeled mask array layouts shown in later sections. For the conventional Lyot coronagraphs, the IWA listed is simply the occulting spot radius, and the listed OWA is the full, unvignetted FOV of the instrument.

| Mask configuration class | Coro. type | Configuration name | Band number | SPAM position | FPAM position | LSAM position | FSAM position | Azim. FOV (deg) | IWA ($\lambda/D$) | IWA (mas) | OWA ($\lambda/D$) | OWA (mas) |
|---|---|---|---|---|---|---|---|---|---|---|---|---|
| Unsupported (low contrast) | Lyot | 0.281" ⌀ occulter | 1, 2 | OPEN | HLC12_C2R1:9, HLC12_C3R1:8, HLC12_C4R7:9 | NFOV, WFOV | OPEN, any | 360 | — | 140.5 | — | 3600 |
| | Lyot | 0.322" ⌀ occulter | 1, 2 | OPEN | HLC12_C4R1:6, HLC12_C5R1:8 | NFOV, WFOV | OPEN, any | 360 | — | 161.1 | — | 3600 |
| | Lyot | 0.803" ⌀ occulter | 1, 2 | OPEN | HLC12_C1R3:4 | NFOV, WFOV | OPEN, any | 360 | — | 401.5 | — | 3600 |
| | Lyot | 0.921" ⌀ occulter | 1, 2 | OPEN | HLC12_C1R2 | NFOV, WFOV | OPEN, any | 360 | — | 460.7 | — | 3600 |
| | Lyot | 1.807" ⌀ occulter | 1, 2 | OPEN | HLC12_C1R1 | NFOV, WFOV | OPEN, any | 360 | — | 903.4 | — | 3600 |
| | Lyot | 0.357" ⌀ occulter | 2, 3, 4 | OPEN | HLC34_R5C1:3, HLC34_R6C1:3, HLC34_R7C1:2, HLC34_R8C1 | NFOV, WFOV | OPEN, any | 360 | — | 178.6 | — | 3600 |
| | Lyot | 0.403" ⌀ occulter | 2, 3, 4 | OPEN | HLC34_R2C1:3, HLC34_R3C1:3, HLC34_R4C1:3 | NFOV, WFOV | OPEN, any | 360 | — | 201.6 | — | 3600 |
| | Lyot | 1.020" ⌀ occulter | 2, 3, 4 | OPEN | HLC34_R1C3 | NFOV, WFOV | OPEN, any | 360 | — | 509.8 | — | 3600 |
| | Lyot | 1.152" ⌀ occulter | 2, 3, 4 | OPEN | HLC34_R1C1 | NFOV, WFOV | OPEN, any | 360 | — | 576.1 | — | 3600 |
| | Lyot | 2.294" ⌀ occulter | 2, 3, 4 | OPEN | HLC34_R1C2 | NFOV, WFOV | OPEN, any | 360 | — | 1147.1 | — | 3600 |
| Unsupported (calibration) | ZWFS | Dual-path ZWFS | 1 | OPEN | HLC12_C7R2:8 | OPEN | OPEN | N/A | N/A | N/A | N/A | N/A |
| | ZWFS | Reflective-only ZWFS | N/A | OPEN | SPC12_R2C2:4, SPC34_R4C1, SPC34_R6C2, SPC34_R8C2 | N/A | N/A | N/A | N/A | N/A | N/A | N/A |

The rows of Table 2 also correspond to those of Fig. 3 except for some additions. For completeness in Table 2, we include four rows for each of the high-contrast HLC occulters because they can also be used at low contrast. In the unsupported calibration section, the table has another row for a fully reflective ZWFS option that blocks all light going to EXCAM and instead sends light at all spatial frequencies to LOCAM.

The fourth column of Tables 1 and 2 specifies the bandpasses in which each mask configuration can work. Only the full bandpasses are listed, but each mask configuration also works in all the subbands of the specified full bands. All of the high-contrast mask configurations work only in their designed bandpasses. The unsupported, low-contrast Lyot coronagraphs are much more flexible and are thus spectrally limited only by the AR coatings used on the glass FPM substrates. The unsupported ZWFS mask configurations work in reflection with LOWFSC, which has its own 128-nm-wide filter centered at 575 nm. The ZWFS that also works in transmission is designed to work in band 1 only and uses in other bands have not been studied.

Columns 5 to 8 of Tables 1 and 2 specify PAM mask positions that combine to form that mask configuration. There is only one FPM design per mask configuration; the multiple FPAM locations listed per row are all copies of that given FPM design, which provides redundancy against manufacturing defects, damage, or contamination. There is only one SPAM mask and one LSAM mask per mode due to space constraints on the PAMs.

There are usually several field stops that can work with each mask configuration because the field stops are not used in a diffractive manner to help create the dark hole. Only the field stops designed specifically for each mask configuration are listed in each row of column 8; other desirable pairings are left for the reader to determine based on the field stop specifications provided in Table 3. Of the 31 field stops at the FSAM, 27 are unique. The positions of the four that are duplicates are shown in parentheses in Tables 1 and 2. All the slit-shaped field stops are intended for use with the spectroscopy modules. The two largest circular field stops (having 1.9-arcsec and 3.5-arcsec radii) are for the low-contrast modes with Wollaston prisms to eliminate crosstalk between polarization states. The FOV is specified by the azimuthal coverage, inner working angle (IWA), and outer working angle (OWA). All configurations have 360-deg azimuthal FOV except for the spectroscopy SPCs, which each have $2 \times 65$ deg coverage. The IWA and OWA are the angular separations at which the off-axis throughput is half the maximum value. With the unsupported, low-contrast Lyot coronagraphs, the IWA column instead lists the occulting spot radius and the OWA column gives the instrument's full, unvignetted FOV in imaging mode. For the multi-star imaging mask configurations,[31] the field of view is limited to the $9 \times 9\,\lambda/D$ field stop, but that field stop can be placed anywhere within the annular field of view listed in Table 1.

### 3.3 Other Possible Mask Configurations

Other than the designed mask configurations listed in Sec. 3.2, there are many other possible combinations of masks and filters. Some may serve a useful purpose, while others are highly discouraged due to poor performance. Either way, all of these other possible mask configurations are unsupported.

In most instances, arbitrary mask combinations would perform worse than the designed ones. All the FPMs are on fused silica substrates with AR coatings, which restrict the bandpasses in which they can be used. Inside the intended bandpasses the reflectivity of each substrate surface is <0.5%, but outside it can be >10%. Substrates in FPAM positions HLC12 and SPC12 can be used in bands 1 and 2, and substrates in FPAM positions HLC34 and SPC34 can be used in bands 2, 3, and 4.

The shaped pupil masks (SPMs) were optimized to make dark holes in their specified bandpasses and fields of view, so decent high-contrast dark holes are not possible out of band or with a different FPM. Because the SPMs increase the brightness of speckles outside the dark hole, one would most likely be better off using the SPAM flat mirror instead of any SPM when performing low-contrast imaging over the instrument's full field of view.

Table 3 Field stops specifications. Mask array positions correspond to those shown in Fig. 15. The actual widths and heights differ slightly from the designed values listed here because of fabrication tolerances as well as observatory and instrument magnification tolerances.

| FSAM position | Primary use | Shape | Width (μm) | Width (λ/D) | Width (mas) | Height (μm) | Height (λ/D) | Height (mas) | Fillet radii (micron) |
|---|---|---|---|---|---|---|---|---|---|
| OPEN | Open | Circle | 7000 | N/A | 14,700 | 7000 | N/A | 14,700 | N/A |
| ND475 | Neutral density (ND) filter | Circle | — | — | — | — | — | — | N/A |
| OPEN2 | Open | Circle | 7000 | N/A | 14,700 | 7000 | N/A | 14,700 | N/A |
| R1C1 | HLC band 1 | Circle | 464.3 | 19.4 | 974 | 464.3 | 19.4 | 974 | N/A |
| R1C2 | SPC major-FWHM vertical slit band 3; tall | Rect | 60.8 | 2 | 127 | 364.8 | 12 | 765 | 30.4 |
| R1C3 | HLC band 1 [spare] | Circle | 464.3 | 19.4 | 974 | 464.3 | 19.4 | 974 | N/A |
| R1C4 | Multi-star imaging in band 4 | Square | 309.1 | 9 | 648 | 309.1 | 9 | 648 | 30 |
| R1C5 | 3.5″ radius circular polarimetry field stop | Circle | 3338.3 | N/A | 7000 | 3338.3 | N/A | 7000 | N/A |
| R2C1 | Multi-star imaging in band 1 | Square | 215.4 | 9 | 452 | 215.4 | 9 | 452 | 30 |
| R2C2 | SPC minor-FWHM rotated slit band 3; tall | Rect | 30.4 | 1 | 64 | 516.6 | 17 | 1083 | 15.2 |
| R2C3 | HLC FWHM vertical slit band 2 | Rect | 30.2 | 1.1 | 63 | 164.8 | 6 | 346 | 15.1 |
| R2C4 | Long slit #3 | Rect | 171.7 | N/A | 360 | 476.9 | N/A | 1000 | 30 |
| R2C5 | SPC major-FWHM vertical slit band 2; short | Rect | 55.0 | 2 | 115 | 164.8 | 6 | 346 | 27.5 |
| R3C1 | SPC ~1st null vertical slit; tall | Rect | 115.7 | 4 | 243 | 347.2 | 12 | 728 | 30 |
| R3C2 | SPC ~1st null vertical slit; short | Rect | 115.7 | 4 | 243 | 173.6 | 6 | 364 | 30 |
| R3C3 | HLC band 2 | Circle | 533.0 | 19.4 | 1118 | 533.0 | 19.4 | 1118 | N/A |
| R3C4 | HLC band 3 | Circle | 589.5 | 19.4 | 1236 | 589.5 | 19.4 | 1236 | N/A |

Table 3 (*Continued*).

| FSAM position | Primary use | Shape | Width (μm) | Width (λ/D) | Width (mas) | Height (μm) | Height (λ/D) | Height (mas) | Fillet radii (micron) |
|---|---|---|---|---|---|---|---|---|---|
| R3C5 | HLC band 4 | Circle | 666.2 | 19.4 | 1397 | 666.2 | 19.4 | 1397 | N/A |
| R4C1 | Multi-star imaging in band 4 [spare] | Square | 309.1 | 9 | 648 | 309.1 | 9 | 648 | 30 |
| R4C2 | SPC minor-FWHM rotated slit band 3; short | Rect | 30.4 | 1 | 64 | 182.3 | 6 | 382 | 15.2 |
| R4C3 | SPC minor-FWHM rotated slit band 2; tall | Rect | 27.5 | 1 | 58 | 467.0 | 17 | 979 | 13.7 |
| R4C4 | SPC minor-FWHM rotated slit band 2; short | Rect | 27.5 | 1 | 58 | 164.8 | 6 | 346 | 13.7 |
| R4C5 | HLC FWHM vertical slit band 3 | Rect | 33.4 | 1.1 | 70 | 182.3 | 6 | 382 | 16.7 |
| R4C6 | SPC major-FWHM vertical slit band 3; short | Rect | 60.8 | 2 | 127 | 182.3 | 6 | 382 | 30.4 |
| R5C1 | HLC FWHM curved slit band 3 | Arc | 33.4 | 1.1 | 70 | N/A | ±45° | N/A | 16.7 |
| R5C2 | HLC FWHM curved slit band 3 (reversed) | Arc | 33.4 | 1.1 | 70 | N/A | ±45° | N/A | 16.7 |
| R5C3 | Left half HLC band 1 | Arc | 232.2 | 9.7 | 487 | 464.3 | 19.4 | 974 | 30 |
| R5C4 | Right half HLC band 1 | Arc | 232.2 | 9.7 | 487 | 464.4 | 19.4 | 974 | 30 |
| R5C5 | empty/dark | N/A | 0 | 0 | 0 | 0 | 0 | 0 | N/A |
| R6C1 | 1.9" radius circular polarimetry field stop | Circle | 1812.2 | N/A | 3800 | 1812.2 | N/A | 3800 | N/A |
| R6C2 | SPC wide FOV halves bands 1 and 4 | Puzzle | 1024.4 | 42.8 | 2148 | 1469.8 | 42.8 | 3082 | 60 |
| R6C3 | Long slit #2 | Rect | 143.1 | N/A | 300 | 476.9 | N/A | 1000 | 30 |
| R6C4 | Long slit #1 | Rect | 85.8 | N/A | 180 | 476.9 | N/A | 1000 | 30 |
| R6C5 | SPC major-FWHM vertical slit band 2; tall | Rect | 55.0 | 2 | 115 | 329.7 | 12 | 691 | 27.5 |
| R6C6 | HLC band 1 [spare] | Circle | 464.3 | 19.4 | 974 | 464.3 | 19.4 | 974 | N/A |

## 4 Design Decisions Common to All Mask Devices

Before delving into the specific masks and design decisions for the mask-holding PAMs, we first discuss design decisions relevant to all those cases.

### 4.1 Mask Layout Constraints

In this paper, we try to present the layout design decisions as a clear, linear storyline when possible. In reality, the process of arranging the masks on each PAM was iterative and somewhat messy. Early on in the project (Phase A and before), the different sub-teams pertinent to the mask layout (e.g., the teams for PAMs, static optics, and mask design) operated mostly independently of one another and used unofficial, notional values to proceed with the initial instrument design. This is because requirements were not made explicitly for how to arrange the various coronagraphic masks, so there was no explicit ownership of this task assigned to a specific team in the project. This gap did not become apparent until the mask designs were being finalized late in Phase B and mask layouts were needed, at which point the mismatching assumptions came to the fore. Fortunately, all the required and contributed masks that were desired were still able to fit onto all the PAMs. The downside was just that the real estate on some PAMs, particularly FPAM, was underutilized, leading to tighter mask packing and fewer copies of masks for redundancy.

By the time late in Phase B the mask configurations were chosen and mask layouts needed to be designed for all the PAMs, nearly all the other constraints in the instrument were already finalized. That includes details such as PAM travel ranges, the outer dimensions of the mask substrates, and a lack of baffling around flexure mounts. The only major mechanical specifications that got to be co-designed with the mask layouts were the baffle size and shape at each mask-holding PAM, which was crucial for packing masks as closely together as possible while still mitigating stray light from neighboring openings.

### 4.2 Fiducial Markings

It was necessary to add fiducials to every individual mask device to tell its correct mounting orientation, both in clocking and in determining the front side. For utmost clarity, one or two arrows were etched on the top of the front face of each mask device to point out which direction should be up when assembling the optical bench. In addition, several $F$-shaped fiducials were added on each mask device in locations that did not interfere with coronagraphic performance—usually along the device perimeter. The $F$ fiducials served two purposes. First, they provided straight lines along both Cartesian axes to use as references when the masks were being aligned by hand under a microscope to their optical interface plates before bonding and again when they were measured post-bonding. When possible, copies of the $F$ fiducials were placed close to each other so they could lie in the same frame of microscope images and also on opposite sides of a substrate to get a longer baseline. Second, they were used to confirm that the masks were in the correct orientation at that plane as specified by the project's parity tracking document, called the Pointing, Positioning, Phasing and Coordinate Systems (PPPCS). For that reason, the $F$ fiducial is rotated or flipped on each device to match the PPPCS at that plane.

Except for a few places on FPAM where the fiducials are at large separations, the arrow and $F$ fiducials are not actually visible in situ using EXCAM. This was intentional as the fiducials are mostly needed during instrument assembly but not testing and can introduce stray light or unwanted diffraction if situated too close to a mask being used.

To calibrate the clocking and apparent magnification of the masks in situ using only EXCAM images, we had to rely on either extra fiducials added at SPAM or use some of the masks themselves as fiducials at the other three mask planes. The extra calibration features at SPAM are shown and described briefly in Sec. 5.3. The magnification and clocking calibrations using fiducials and masks will be detailed in a future publication on alignment and calibration for the Roman Coronagraph Instrument.

### 4.3 Counteracting Mask Foreshortening

All of the mask substrates have a nonzero angle of incidence with respect to the chief ray: 7.5 deg at SPAM, 5.5 deg at FPAM, 7.6 deg at LSAM, and 5.0 deg at FSAM. This causes an apparent foreshortening of each mask that we must counteract for the high-contrast designs to work as intended. Each mask was therefore elongated along the horizontal axis of the instrument's optical bench to exactly counteract the foreshortening. As an example, the circular field stops are actually elliptical holes but appear as circles on EXCAM. Although not mentioned again, this foreshortening compensation was applied to every individual coronagraphic mask but not to the substrate dimensions, inter-mask spacings, or fiducial markings.

## 5 SPAM Masks

The first mask-carrying PAM along the optical train is SPAM, which receives a 17.0-mm-diameter beam conjugate to the input pupil. As shown in Fig. 4, SPAM carries four different SPMs for the SPCs as well as a standard fold mirror used for the HLCs and calibrations. The official names of the positions are OPEN for the fold mirror, WFOV for the wide FOV SPM, SPECROT for the rotated spectroscopy SPM, SPEC for the primary spectroscopy SPM, and MSWC for the MSWC SPM. All four masks are manufactured together on a 44-mm-wide by 47-mm-high by 4-mm-thick silicon substrate provided by the Japan Aerospace Exploration Agency (JAXA). The black regions on the substrate front face are cryogenic black silicon developed at JPL's Microdevices Laboratory (MDL), and the gray areas are bare aluminum.[36–38] The small, 1-mm-diameter circles are used for calibration only and can only be seen when the beam is centered halfway between either the top row or bottom row of the SPMs.

### 5.1 SPAM Layout Design Constraints

The largest decision for the SPAM layout was whether to carry one or two optics. The deciding factor was that the required HLC would have significantly higher throughput if the SPAM open position were its own optic. Because of the flexures needed around the standalone fold mirror, that leaves enough room for a $2 \times 2$ grid of masks on another substrate within the PAM travel range. The chosen fold mirror is a standard flat mirror with a protected silver coating for the highest reflectivity (>97%) in the instrument's operating bands. The SPMs, on the other hand, are made with lower reflectance (85% to 91%) bare aluminum because protected silver cannot survive the high and cryogenic temperatures during the SPM fabrication.

Despite the silicon substrate for the SPMs being as large as it could be—44.0-mm-wide by 47.0-mm high by 4.0-mm thick—for the underlying optical interface plate (OIP), it was difficult to fit in all the desired aspects of the SPM mask array. To be able to pick up the device with tweezers without incurring damage, it was decided that all the black silicon must be at least 2.0 mm from the sides of the device. The next constraint was that the usable horizontal travel (the same as the horizontal axis in Fig. 4) was restricted to within 30.6 mm of the left side of the silicon wafer. This leaves less than 100 microns of travel margin after taking into account the possible range of instrument and observatory alignment tolerances. With the masks pushed as far left on the device as they could to be within the travel range, and aiming for the same amount of black silicon buffer for stray light blocking around each mask, the amount of buffer on each side ended up being ∼1.6 mm. The vertical travel constraints resulted in the same amount of black buffer around each mask in that axis as well. The SPAM baffle, located 14.5-mm upstream of the SPAM reflective surfaces, was chosen to be a racetrack shape to allow the full 17.0-mm-diameter beam to pass through before and after reflection with ∼1.0 mm of clearance all around. When combined with alignment tolerances, the ≈1.6 mm of black silicon buffer around each SPM should be enough to block any stray light outside the pupil except on the left side (as oriented in Fig. 4) because of the racetrack shape. This should be a small risk, though. Unlike in a typical coronagraphic testbed where an OAP is generally overfilled and stopped down to define the pupil mask, any light outside the pupil at SPAM would be incoherent with the main beam because there is no mirror outside the primary. Then, any incoherent light there would only be mildly affected by the FPM and then be blocked by Lyot stop at LSAM. One might notice that there was room for another 1.4 mm of black silicon on the rightmost side of the substrate, but that would be unnecessary as the right side is fully blocked by the racetrack baffle.

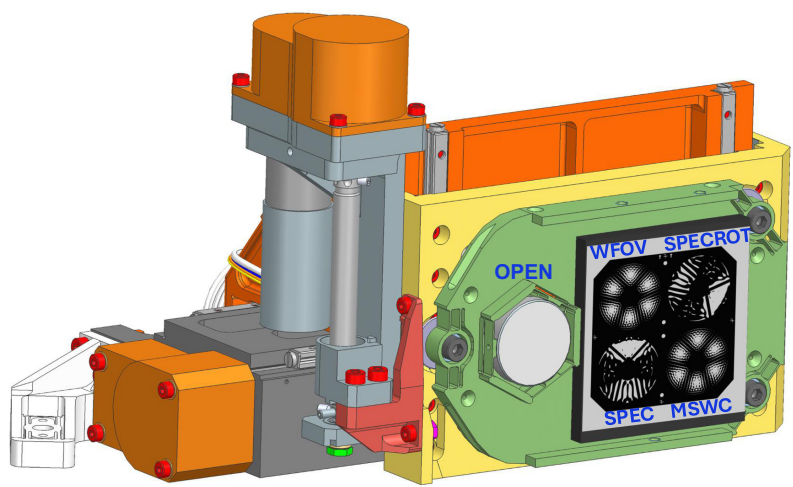

**Fig. 4** Shaped pupil alignment mechanism (SPAM) in the Roman Coronagraph Instrument. The five nominal positions are labeled. The OPEN position is for calibrations and the HLC, and the four mask positions to the right are for the SPCs.

### 5.2 Mask Choices and Redundancy

Once it was certain we would have four mask positions at SPAM, we had to choose which SPM designs to use and where to put them in the 2 × 2 grid. Dating back to the ACWG final decision in 2013,[29] there were two definite SPC modes chosen for the Roman Coronagraph Instrument. One is the wide FOV option to provide nearly the largest DM-controllable FOV ($20.1\,\lambda/D$) at the expense of a large IWA as well ($5.9\,\lambda/D$). The other was spectroscopy mode with a smaller ($3\,\lambda/D$) IWA at the expense of a bowtie-shaped dark hole covering only about a third of the possible azimuthal extent. Because of concern over the aforementioned tiny horizontal stage travel margin for the right column, the two ACWG-selected, best-effort SPMs for spectroscopy and wide FOV were placed in the left column. As for which mask should go in the top or bottom row, that was an arbitrary choice because the travel range margin was the same in the bench-vertical direction.

For the remaining two SPM positions, there were only three options in contention. The two most obvious options were the two other orientations, rotated 60 deg and 120 deg, of the spectroscopy dark hole that had been part of the original design. In 2020, a working group studying the science capable with two versus all three orientations found that there is only a 2% to 3% reduction in the possible observing efficiency of the prime target exoplanet 47 Uma c. The telescope can roll ±13 deg combined with waiting a few months allowing almost all the rest of the azimuthal coverage to be reached.

The third, newer option being considered was for MSWC. There is strong scientific interest in this technique because it could allow us to image around nearby binary systems, most intriguingly alpha Centauri A and B. Although at a lower technology readiness level (TRL) than the third spectroscopy orientation, the MSWC mask was deemed a higher priority because of its higher scientific potential. The key apodizer change for MSWC is to put a grid of spots in the pupil plane to create the high-frequency diffraction orders around which a super-Nyquist dark hole can be dug for the off-axis star. That still left the underlying mask to be decided. One key deciding factor was that there were no more Lyot stop positions left (as discussed later in Sec. 7), so the MSWC coronagraph design would have to re-use a Lyot stop, and therefore probably be very similar to one of the existing other three modes. The MSWC concept relies on dark holes for the on-axis and off-axis stars at different spatial frequencies to overlap on the detector, which essentially ruled out the two spectroscopy SPCs because of their small bowtie-shaped dark holes. In terms of ease of doing MSWC, the HLC would be more versatile and amenable because there are many viable DM shapes that can work well together. This is in contrast to the SPC, for which the on-axis star wants a flat wavefront for its dark hole and the off-axis star wants a larger WFE to dig down the pinned speckles near it. However, the prime scientific case for MSWC with the Roman Coronagraph Instrument is alpha Cen. Because alpha Cen is so close, the stellar diameters are huge compared with other nearby stars (6 and 8 mas versus 1 mas). The HLC, with its smaller IWA, is much more sensitive to tip/tilt and therefore sees orders of magnitude worse contrast than the SPC-wide FOV. In addition, the habitable zone around alpha Cen goes out to about an arcsec, which is also about the FOV for the on-axis wide-FOV mode. Therefore, it was an easy decision that the MSWC pupil mask should be a near-copy of the wide FOV SPM but with a grid of spots integrated into it. The numerical optimization was actually re-run to include those spots so that the nominal PSF for just the on-axis star still gives the same $10^{-9}$ contrast dark hole as the wide FOV SPM in a perfect optical system, just with a slightly lower throughput because of the opaque spots in the pupil.

For redundancy, this worked out well with two spectroscopy SPMs (in two orientations) and two WFOV SPMs (one with the grid of spots included in it). To be extra safe in case one row of the SPMs was accidentally damaged or was unreachable with the bench-vertical SPAM stage, the similar use case SPMs were put diagonal to each other in the $2 \times 2$ grid.

As described by Gersh-Range et al.,[24] the orientation chosen for the rotated spectroscopy SPM was for the pupil rotated by 60 deg. This orientation provided slightly higher throughput than when the pupil was rotated 120 deg. The SPC optimization was numerically complex enough that one axis of symmetry had to be enforced in the apodizer plane for the problem to be tractable. The Roman Coronagraph Instrument's input pupil has a near-mirror symmetry about only one axis, the bench vertical (because the instrument itself is 0.4 deg off-axis), so the hexapod strut structure becomes more asymmetric for the other possible rotations of 60 deg and 120 deg.

### 5.3 Calibration Circles

A last-minute but ultimately critical addition to the SPAM mask array was two pairs of 1.000-mm-diameter (1/17 beam diameter) circles centered 15.000 mm apart and halfway between the columns of the apodizers. We decided that for alignment and calibration of the SPAM mask array, for which the algorithms and software had not been planned yet, it would be much easier to identify and isolate small circles in the image than the undersampled and partially defocused pupil images of the apodizers themselves. (By undersampled, we mean that the design grid of pixels in the apodizers is $1000 \times 1000$, but the beam diameter in the EXCAM pupil imaging mode is only 300 pixels across.) In the end, this was a quite helpful decision. As shown in Fig. 5, the combination of the two small circles far apart in the same image was useful for determining the magnification and clocking of the SPAM mask array. Because the diameter and separation of the circles is a fixed ratio, even poor imaging quality can only affect the edges of the circles but not their separation. In addition, for calibrating the apparent motion of SPAM as seen on EXCAM, we were able to isolate just one circle in the frame and move it several millimeters along each axis as shown in Fig. 6. We then utilized standard circle detection algorithms in Python libraries to determine the mask array motion.

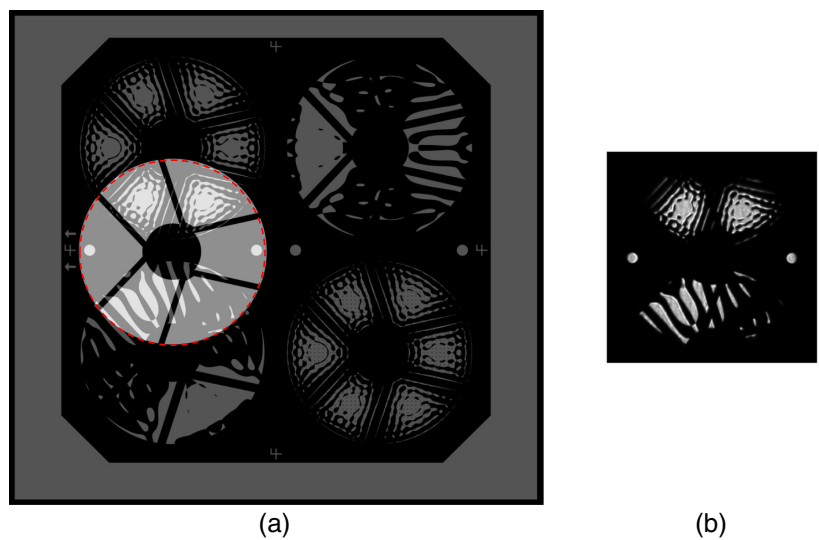

(a)         (b)

**Fig. 5** SPAM position for magnification and rotation calibration as oriented on EXCAM. (a) The beam footprint, shown partially transparent and circled, is placed halfway between the WFOV and SPECROT positions to illuminate the two fiducial circles on the far left and far right of the beam. (b) The image taken during TVAC testing of the instrument was used to compute the magnification and clocking of the masks at SPAM relative to the model-based, expected values.

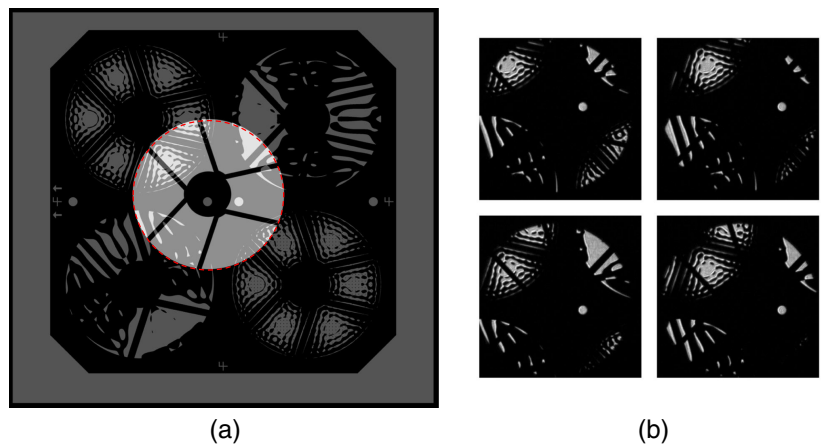

(a)          (b)

**Fig. 6** SPAM positions for motion calibration as oriented on EXCAM. (a) The beam footprint, shown partially transparent and circled, is placed slightly off-center from the SPAM mask array to leave unblocked just one of the circular fiducials. (b) EXCAM images taken during in-air testing of the instrument for four slightly different SPAM positions. Tracking the circle position allows for simple calculation of the $2 \times 2$ matrix between SPAM and EXCAM coordinates.

## 6 FPAM Masks

After SPAM, the next mask-holding PAM in the optical train is FPAM. It was by far the most difficult PAM to lay out with four separate mask substrates and dozens of masks in total.

### 6.1 Layout Design Constraints for FPAM

There are two independent ways to organize the FPMs, resulting in there being four FPAM mask substrates. The first split is between masks that are obscurations with clear surroundings (i.e., HLC and conventional Lyot occulters) and masks that are openings with opaque surroundings (i.e., SPC FPMs and pinholes). That fundamental difference between mask surroundings needing to be fully clear or fully blocked meant that the HLC and SPC FPMs could not fit on the same 23-mm-diameter substrates. The second division is based on bandpass. The two AR coatings used for the fused silica flight substrates were for bands 1 and 2 or for bands 2, 3, and 4. Because there are an equal number (6) of high-contrast mask configurations in bands 1 and 2 and bands 3 and 4, the masks were also grouped in that way. This led to the names of the FPAM mask substrates being HLC12, HLC34, SPC12, and SPC34 as shown in Fig. 7. The one exception already mentioned was that the simpler, replacement band 2 HLC occulters with a larger pointing control capture range were placed on the contributed HLC34 substrate because the required HLC12 substrate had already been manufactured and accepted by that time in spring 2021. Besides the masks, the other three main positions on FPAM are a through-hole position named HOLE for taking LOCAM dark frames and two neutral density (ND) filters with optical densities of 2.25 and 4.75 named ND225 and ND475 for unocculted PSF observations.

    The next mask layout design constraints for FPAM are related to the stage travel ranges. As shown by the pink rectangle in Fig. 7, the allowable FPAM travel range omits one or two edges on each of the four mask substrates. A lack of requirements on mask layouts caused different assumptions between teams that were not caught until right before the mask layouts had to be finalized. The notional mask layouts carried by the optomechanical team had only seven masks in a one-ring hexagonal layout with 4.5-mm spacing and centered on each substrate, but the mask design team had assumed that the whole substrate surfaces were usable. The PAM specifications were already frozen, so the mask layouts were designed around the given travel range restrictions.

    The next big question was where each of the four substrates should be placed. Figure 7 shows that the two middle mask substrates have no horizontal limitations, but the two corner

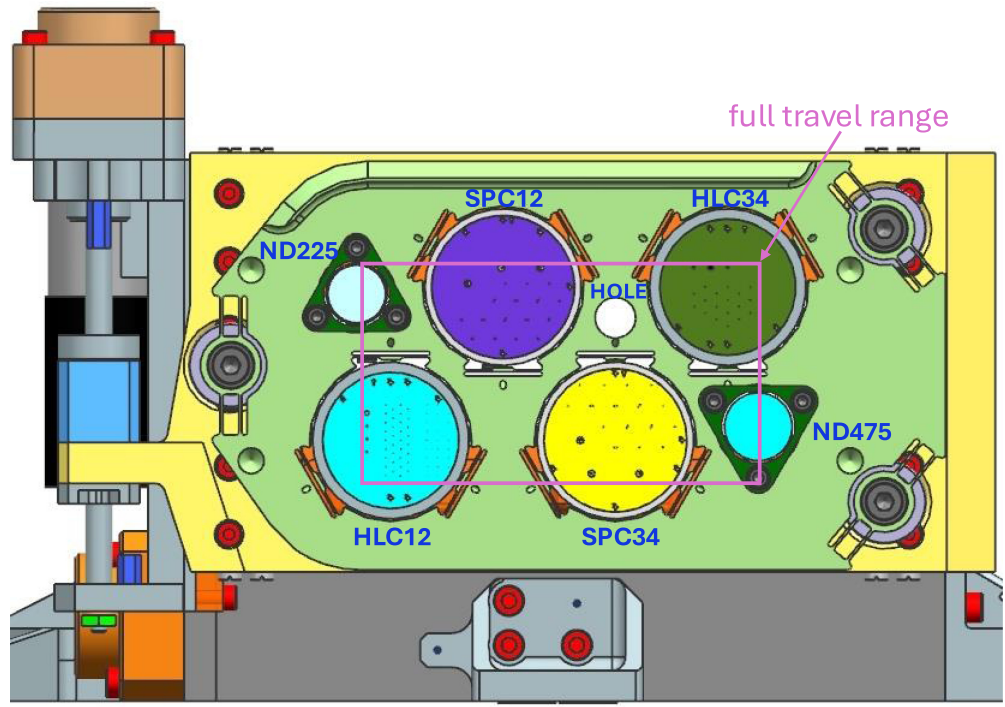

**Fig. 7** Focal plane mask alignment mechanism (FPAM) in the Roman Coronagraph Instrument. The seven positions are labeled: four for HLC or SPC masks in either bands 1 and 2 or 3 and 4, a through-hole, and two neutral density filters with optical densities of 2.25 and 4.75. A rectangle shows the full possible travel range in terms of where the PSF can land. Some regions of each mask substrate are not accessible.

positions have some unreachable areas both vertically and horizontally. To determine which FPMs needed more spacing than others, we performed some open-loop experiments in a simple Fourier numerical model of contrast degradation versus FPM proximity to neighboring FPMs and substrate edges.

The size and shape of the FPAM baffle, shown later in Fig. 11, had to be co-designed with the mask layouts to allow as much of the full field of view as possible but also not allow any stray light through the rough areas past the clear apertures of the substrate nor through the through-hole surrounding the substrate due to the flexure mounts. The instrument's FOV in imaging mode, before accounting for the four mask PAM baffles, is unvignetted out to almost 4 arcsec and then falls off roughly linearly out to a radius of about 8 arcsec. To preserve most of that FOV but not oversize the baffle too much to fit in more FPMs, we chose a minimum baffle FOV of 6.25″ with an additional 2 arcsec tolerance for the on-orbit line of sight choice and 0.40-mm tolerance for lateral baffle alignment. This gives a total baffle radius of 3.73 mm. There is also a small bulge off to the left side of the baffle (as viewed from the front) to fully accommodate the smaller FOV needed in reflection of the LOBE.

The HLC occulters, having relatively small areal coverage with diameters of $5.6\,\lambda/D$, can be packed into a tight grid. Our calculations showed a $1.3 \times 10^{-9}$ contrast degradation in open loop (but none in closed loop) for occulters spaced $50\,\lambda/D$ apart center-to-center in a hexagonal grid, $2 \times 10^{-10}$ contrast degradation for $60\,\lambda/D$ separation, and $1 \times 10^{-10}$ contrast degradation for $70\,\lambda/D$ separation. Therefore, to balance fitting in more masks and not risking affecting the contrast error budget, we opted for $60\,\lambda/D$ (=1.13 mm in band 1) separation for HLC occulters. For the SPC FPM layouts, a major design constraint was that the unpolished outer 1 mm along the substrate edge could not be coated for the risk of the metal flaking off. Because the FPMs are optimized assuming no light outside the mask opening, we had to place all the FPMs far enough away that the outer uncoated edge would not be visible at all through the baffle opening.

The annular wide FOV SPC FPMs have a relatively enormous areal coverage because of their $40.8\,\lambda/D$ outer diameter. We found that within the baffle opening, there was no safe separation distance between wide FOV FPMs. Otherwise, the open-loop contrast would degrade by at least two orders of magnitude. Another reason for not having other openings visible within the baffle is that the wide FOV SPM is also used for the unsupported multi-star mode, so there cannot be any chance of the off-axis star being unocculted.

The bowtie-shaped spectroscopy SPC FPMs, having an intermediate areal coverage because of their $18.8\,\lambda/D$ outer diameter, can tolerate an intermediate spacing distance. We ultimately chose a center-to-center bowtie spacing of $150\,\lambda/D$ as that changed the speckle morphology at $\approx 10^{-9}$ contrast but did not make the average contrast worse. We decided that with aberrations and in a closed loop the change in speckle morphology would probably not matter, and it was worth the trade of packing in more masks versus needing a much larger separation to avoid any diffraction effects from neighboring openings.

Finally, we had to determine the spacings for the $0.5\,\lambda/D$ diameter pinholes included for isolating aberrations downstream of FPAM during phase retrieval calibrations. Because of the dynamic range of the detector, we determined that any light from the PSF past $25\,\lambda/D$ would be lost in the noise. This allowed the pinholes to be placed closer to the substrate edges than the SPC FPMs.

Given those constraints, we could start to narrow down the placement of each of the four FPAM mask substrates. The lower two positions as shown in Fig. 7 have a larger vertical travel range, so they were designated for the required and best-effort FPMs (except for the band 1 wide FOV FPMs that were not considered best-effort until years later in TVAC). Because the SPC FPMs need larger separations and the required HLC FPMs could all easily fit on the lower left position, the SPC34 substrate was given the most travel range in the lower right position. The upper two positions remained for the rest of the FPMs which were all unsupported at the time. Again, the SPC FPMs need larger separations, so the SPC12 substrate was assigned the upper left position for its full lateral travel range.

### 6.2 Mask Choices and Redundancy

With the positions of the four mask substrates decided, the next phase of the layout design was to determine how many of each FPM to include and where. This involved many hours of iterating on diagrams with hexagonal mask grids, the outline of the baffle transmission, and row versus column major grid layouts.

There were no requirements on how many instances of each mask design to put on the FPAM mask substrates, so we opted to put as many feasible of each. We had assumed a worst-case fabrication yield of 50% per FPM design on a given substrate and then wanted another backup in case of contamination or damage after fabrication, so we desired a minimum count of three of each FPM design. Beyond that, the other major consideration was particulate contamination during launch, when all contaminants still within the instrument were assumed to be redistributed evenly. We quickly found that different estimates for the instrument cleanliness could result in the need for between a handful and hundreds of FPMs to have a high likelihood of having one with no contamination in the dark-hole area. The only way to know on orbit if dust is contaminating the FPM is to perform HOWFSC with it, but realistically, there is only enough time on orbit to do that for a few (e.g., five) masks. On the HLC12 substrate, however, we did place more FPMs than that as the dielectric layer thickness can vary across the substrate and we wanted to be able to choose the one most closely matching the design. Because of the large spacing needed between SPC FPMs, we were able to fit only three of each type onto each substrate.

Several necessary calibration positions had to be spread across the substrates. As stated before, the HLC substrates are mostly clear with opaque masks whereas the SPC substrates are mostly opaque with clear mask openings. The open position for all of bands 1 and 2 is thus on the HLC12 substrate and likewise for bands 3 and 4 using the HLC34 substrate. This is not ideal for the SPCs because of beam shear when switching between substrates, but it is the only viable option. The pinholes for characterizing the phase downstream of FPAM during TVAC needed large opaque regions around them, so those were placed on substrates SPC12 and SPC34. (Note that due to time constraints during TVAC, the downstream aberrations were only able to be measured in band 1. On-orbit characterization will not be feasible because the pointing control loop cannot be running while the pinhole is in use, and the $\approx 10$ mas observatory jitter will be too large for the star to couple into the pinhole. Fortunately, the downstream aberrations were measured to only be a few nm RMS in band 1 during TVAC and could either be scaled or ignored completely in the other bands with little to no effect on HOWFSC performance.)

### 6.2.1 Layout of the FPAM substrate HLC12

The HLC12 substrate diagrammed in Fig. 8 is the most important as it carries the band 1 HLC FPM required for TTR5. The first constraint considered was what to put along the left edge of the travel range limit on the substrate as shown in Fig. 7. The four large conventional Lyot occulters and the open/clear position for calibrations want the largest clear FOV possible, so those were a natural fit in a column along that travel range edge. There was not enough room on the substrate to place the large occulters completely out of each other's FOV, so it was accepted that they would block some of each other's FOV along the vertical axis. The open position was placed in the bottom left corner of the travel range so that only one quadrant of its FOV had to use up valuable real estate for the high-contrast masks. This first column forced the rest of the layout to be most easily optimized as column-major as well.

The diameters of the conventional Lyot coronagraphs were chosen to be complementary to the FOVs of the high-contrast mask configurations, which have OWAs of 9.1, 9.7, and 20.4 $\lambda/D$. The radii of the Lyot occulters were chosen as 8.0 and 18.0 $\lambda/D$ in bands 1 and 2 to be as large as possible to block more starlight but still have a little overlap with the high-contrast zones for stitching together full-frame images. Due to time constraints, no fine-tuning or performance simulations were done for these low-contrast coronagraphs.

In the middle of the substrate is the safest area (in terms of stray light and epoxy near the edges) for the high-contrast FPMs for bands 1 and 2. With the chosen $\lambda/D$ FPM spacing, there were 34 positions available. The first 20 in columns 2 and 3 and the bottom of 4 are all the required band 1 designs, and the rest are the alternate band 2 designs that were the primary at the time. In retrospect, 20 and 14 of the two designs were excessive. We should have included some backup, alternative versions of both FPMs that used the traditional, inverted top-hat style ZWFS dimple for easier fabrication and better LOWFSC performance. At the time, however, our plan for HOWFSC did not involve the relinearization of the DM response matrix and the complicated HLC occulter designs were the only ones that could achieve TTR5 in simulation. So by being too strict at the time with an all-or-nothing approach to meeting requirements, we prevented ourselves from including a more robust, simpler design that could have provided similar HOWFSC performance and better LOWFSC performance.

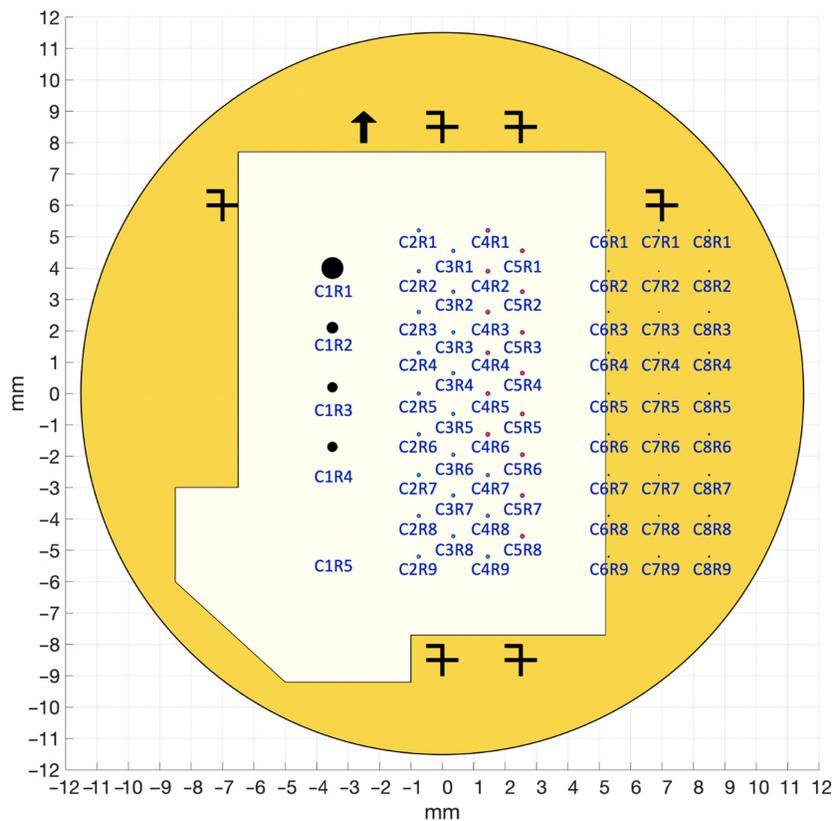

**Fig. 8** HLC12 substrate at FPAM with positions labeled as used in Tables 1 and 2. The outer yellow region in the diagram is where the resist, PMGI, was left on. The plate scale at FPAM is 2.1 arcsec/mm and 17.8 microns per $\lambda_1/D$. This substrate carries the open position for bands 1 and 2 as well as the FPMs for band 1 HLC, alternate band 2 HLC, band 1 or 2 Lyot coronagraphs, and dual-path ZWFS.

The dual-path ZWFS is an unsupported mode on the HLC12 substrate meant to operate simultaneously for differential wavefront sensing on both EXCAM and LOCAM. The dual-path ZWFS is not as sensitive to stray light and could lie closer to the substrate edge than the high-contrast FPMs. These spots have to be in a transmissive medium, which was chosen to be the resist polydimethylglutarimide (PMGI) that was already being used for phase control on the HLC occulters. PMGI was therefore left on the right side of the HLC12 substrate as well as anywhere else outside the FOV of the FPMs because the etching time scales with the amount of resist taken off. Because the ZWFS phase spots are invisible on EXCAM, we sandwiched the seven dual-path ZWFS spot locations (column 7, rows 2 to 8) exactly halfway between metal spots to the left and right. There are also metal spots above and below in rows 1 and 9. The intention is that the star could be aligned to the metal spots on the left and right of a ZWFS spot, and then, the PAM stages could be driven to the middle position. Finally, some as-yet-to-be-written software would perform the fine alignment from there using either EXCAM or LOCAM images. One regret in this plan is that the metal alignment spots were made quite small at just $2\lambda/D$ in diameter to block less of the FOV, but it would have been much better to use the same $5.6\lambda/D$ diameter of the high-contrast FPMs so that existing flight software fine-alignment algorithms could be re-used directly.

### 6.2.2 Layout of the FPAM substrate HLC34

The HLC34 substrate shown in Fig. 9 was laid out with the same strategy as the HLC12 substrate. This time the longer travel range was along the horizontal axis, so the conventional Lyot occulters and open position were placed along the upper boundary in a row. The shorter travel range on this substrate meant there was only room for three conventional Lyot occulters, so they were chosen as $8.0\lambda_3/D$, $8.0\lambda_4/D$, and $8.0\lambda_4/D$; the $18.0\lambda_3/D$ option was the one skipped because there is not a wide FOV option in band 3 anyway. The main open position is in the upper right corner (position R1C4) to waste the least amount of space for masks, and an alternate, slightly less open position was placed at position R8C1 because there was room.

There are no ZWFS phase spots on the HLC34 substrate, so the rest of the positions are for high-contrast FPMs. Originally, there were only HLC occulters for the band 3 and 4 designs, but the LOWFSC study showing poor capture range in bands 2 and 4 necessitated a late redesign to include occulters in those bands that could tolerate the worst-case pointing control capture range scenarios. Row 2 is all the alternate, original band 4 HLC occulter designs with the small pointing control capture range. Below in rows 3 and 4 are the six main band 4 HLC occulter designs with the simple dimple PMGI profiles for better LOWFSC performance. Rows 5 and 6 are band 3 HLC occulters, which work well with LOWFSC as originally designed. Finally, rows 7 and 8 have four total of the main band 2 HLC occulters with the simple dimple PMGI profiles.

### 6.2.3 Layout of the FPAM substrate SPC34

The SPC34 substrate shown with position labels in Fig. 10 carries the SPC FPMs for bands 3 and 4. Because the wide FOV FPMs had to be isolated from all other openings within the FPAM baffle, we started with their placement before fitting the other masks into the rest of the available area. The best use of space was to put the wide FOV FPMs along the vertical travel range limit on the bottom of the SPC34 substrate because the metalized region below it was unusable for anything else. As diagrammed in Fig. 11, we centered an outline of the full possible baffle illumination envelope on each FPM position to ensure adequate separation from both other masks and the transparent outer ring of the substrate in the final layout. (Only five positions have the spacing guide overlaid for clarity.) Only two of the desired three wide FOV FPMs could fit along the bottom border in row 8, so the third position was placed as far left as it could in row 5. At the top and middle of the SPC34 substrate in rows 2, 4, and 6, there was enough remaining room to place three FPMs for each spectroscopy and rotated spectroscopy SPCs. They were placed in alternating positions to avoid losing all of any one type in the unlikely event of localized damage from events such as epoxy spatter or mishandling.

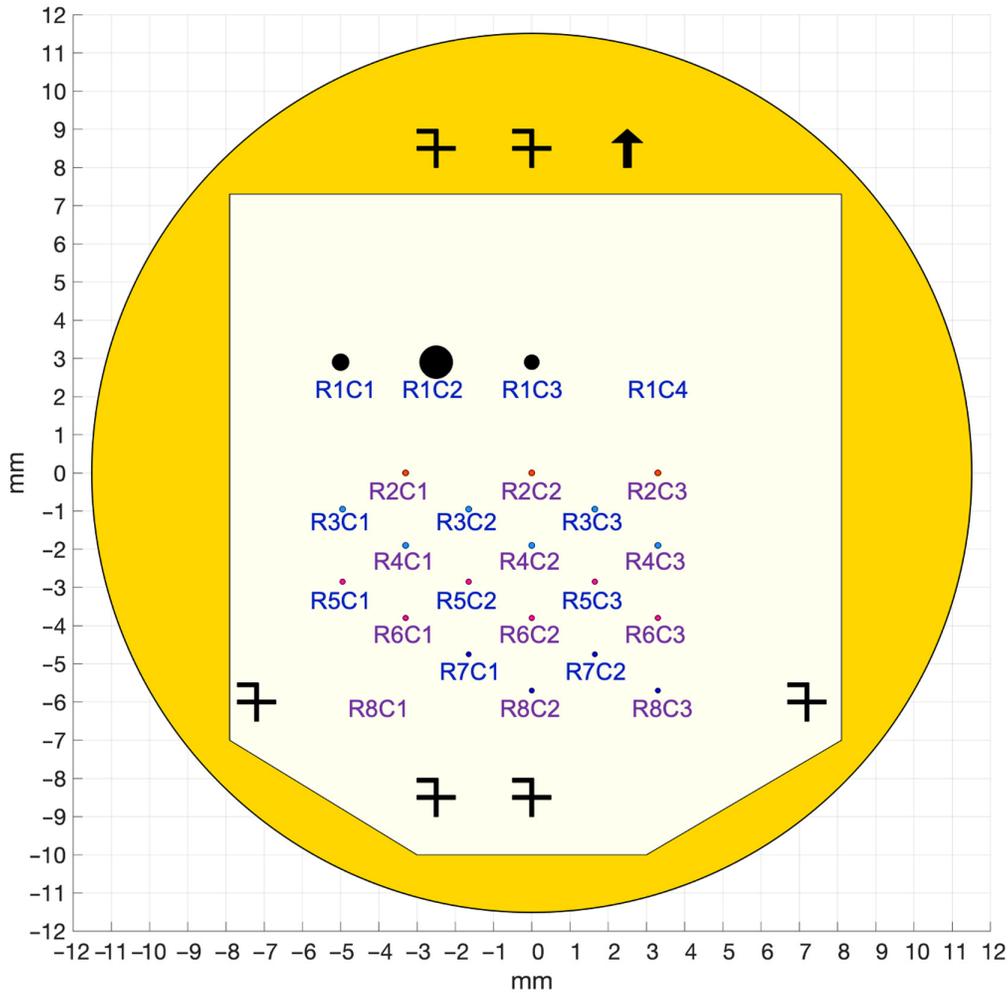

**Fig. 9** HLC34 substrate at FPAM with positions labeled as used in Tables 1 and 2. The outer yellow region in the diagram is where the resist, PMGI, was left on. This substrate carries the open position for bands 3 and 4 as well as the FPMs for the band 2 HLC, band 3 HLC, band 4 HLC, alternate band 4 HLC, and bands 2 to 4 Lyot coronagraphs.

Closer to the edge of the substrate on the three sides reachable by the stages, we placed calibration pinholes because they do not need as much of a lateral buffer. The pinholes sized for $1\lambda/D$ diameter in band 4 were placed in the first three columns of row 1. The pinholes for band 4 were placed along the right side in positions R1C4, R2C5, R6C3, and R7C2. Finally, we placed a backup band 1 pinhole on the left side of the substrate in positions R2C1, R3C1, and R7C1. The band 1 pinholes would experience severe ghosting because of the mismatched band and AR-coating, but they were included because this substrate was more likely to be fabricated as a best-effort device, whereas there was no guarantee that the better-suited but unsupported SPC12 substrate would be fabricated.

In the otherwise unusable vicinity of the three wide FOV FPMs at positions R4C1, R6C2, and R8C2, there are additive spots, or "pimples," for an unsupported, reflective-only LOWFSC capability. The idea is that all spatial frequencies are reflected to LOCAM in this mask configuration, so one could use the images for high-order Zernike wavefront sensing as well, limited only by the low LOCAM sampling. The original sampling on LOCAM was 50 pixels across the pupil, which would have more closely matched the $\approx 47$ DM actuators across the beam and possibly enabled DM drift measurements on LOCAM. However, the final LOCAM sampling is now 38 pixels across, so individual actuators would no longer be able to be resolved.

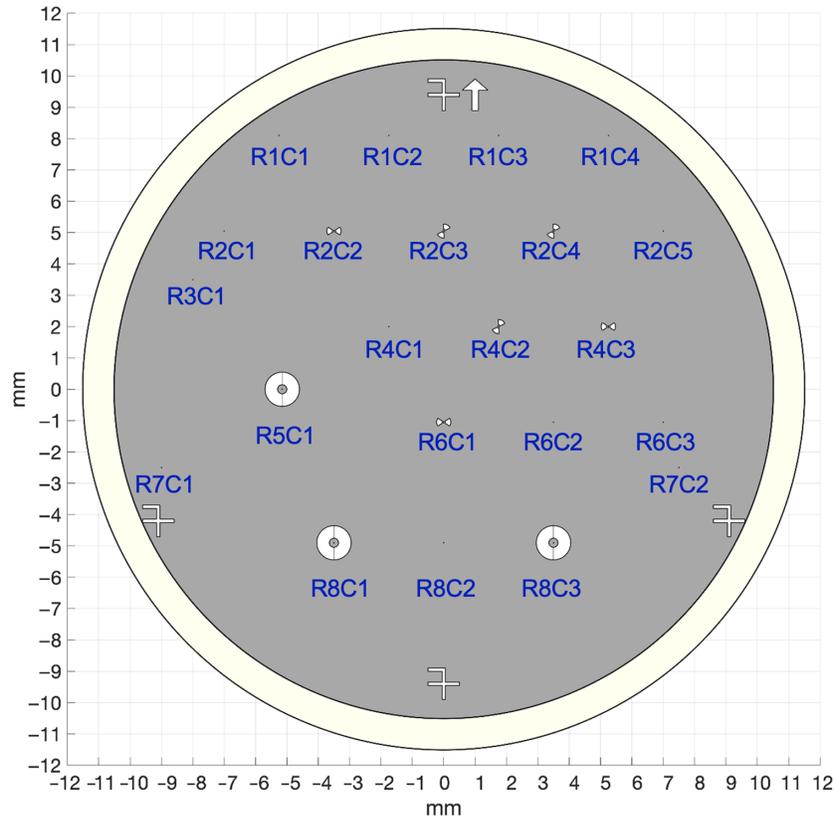

**Fig. 10** Transmissive SPC34 substrate at FPAM with positions labeled as used in Tables 1 and 2. The gray areas are coated with aluminum, and the outer 1 mm in radius is unpolished and uncoated. This substrate carries the spectroscopy and rotated spectroscopy FPMs for SPCs in band 3, the wide FOV FPMs for the SPC in band 4, and calibration pinholes for bands 1, 3, and 4. The vertical line in the annular FPMs is not in the actual device and is a plotting artifact from having to etch the area as two halves to leave the central spot.

### 6.2.4 Layout of the FPAM substrate SPC12

The SPC12 substrate shown with position labels in Fig. 12 carries the SPC FPMs for bands 1 and 2. The design considerations were similar to the SPC34 substrate but flipped vertically because of the travel range restriction now being on the upper side. Therefore, the wide FOV FPMs were placed along the upper edge in row 1 and along the left edge in position R2C1. The spectroscopy and rotated spectroscopy bowtie FPMs fit neatly into row 3 (all columns) and row 4 (columns 2 to 4). Again, the bowtie orientations were alternated. Six $1\,\lambda/D$ pinholes for band 1 wrap around the bottom edge of the substrate in the dead zone too close to the outer clear edge for FPMs. These are positions R4C1, R4C5, and all of row 5. Finally, in row 2 below the wide FOV openings, there are three more unsupported, reflective-only LOWFSC pimples for reflecting all spatial frequencies to LOCAM.

## 7 LSAM Masks

At the next pupil plane after FSAM is LSAM, which holds the Lyot stops. The unmasked beam diameter here is 17.0 mm—the same as at SPAM.

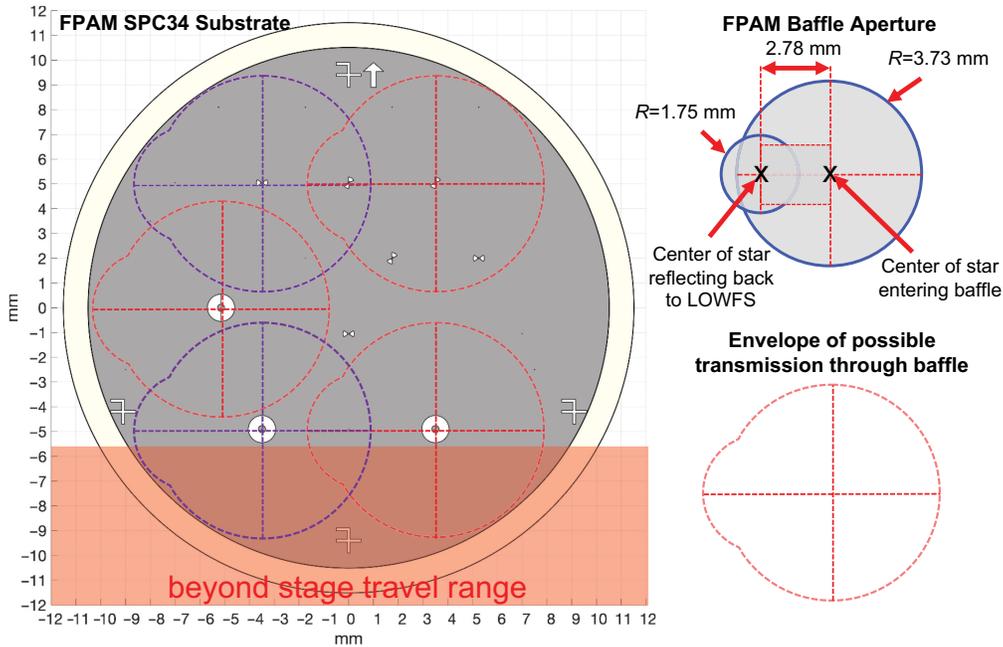

**Fig. 11** Visual guide to how the SPC34 substrate layout was chosen with the mask spacing guide shown at five positions. The FPAM baffle is a circular opening with an extra cutout on the left (as viewed from the front) for the beam reflecting to LOBE. The mask spacing guide is an enlarged version of the baffle to account for baffle misalignment, changes in the chosen line of sight, and the marginal ray from the baffle being 14.5 mm in front of FPAM. All the SPC FPMs are located such that the outer transparent ring of the substrate is fully blocked by the baffle. The wide FOV FPMs are fully isolated because their performance would severely degrade from any other openings being visible at the same time. Another baffle around the mask perimeter would have allowed the placement of more FPMs closer to the edge of the substrate.

### 7.1 LSAM Layout Design Constraints

The layout of LSAM, shown in Fig. 13, is the simplest of the mask PAMs. Five positions were settled on in approximately an *X*-layout based on the number of positions wanted, the necessary inter-mask spacing, and the allowed travel ranges of the stages. This layout of five positions has only a few hundred microns of lateral tolerance past the centers of the corner mask positions. Because a Lyot-style coronagraph diffracts light from the focal plane mask outside the nominal pupil at the Lyot plane, no light from neighboring openings is allowed to pass through in the LSAM mask grid. Therefore, the LSAM baffle, placed 14.5 mm in front of the OIP, is only barely oversized with respect to the beam at 19 mm in diameter except for on one side because of a racetrack opening to allow the reflected beam to bounce unimpeded to a beam dump.

### 7.2 Mask Choices and Redundancy

Because LSAM is in a pupil plane, the large beam diameter precludes the ability to have exact spares of any Lyot stop, at least without eliminating some other capability. Thus, only one of each Lyot stop is carried. The various coronagraphs were designed to have the FPM scale with wavelength but otherwise be as similar as possible, so the same Lyot stop can be re-used for each type of coronagraph in several bandpasses. The position named NFOV is the Lyot stop for the required narrow FOV HLC in band 1 and is also used by the unsupported HLCs in bands 2, 3, and 4. The SPEC position is for the nominal orientation spectroscopy SPCs in bands 2 and 3, and the SPECROT position is for the rotated spectroscopy SPCs in those same bands. Finally, the WFOV position serves as the Lyot stop for both the wide FOV and multi-star SPC designs in bands 1 and 4.

For calibrations of the unmasked beam, there is an unobstructed through-hole at LSAM. To avoid glinting off the relatively thick metal edges of the OIP and because the angle of incidence (AOI) is 7.6 deg at LSAM, the hole is offset at the center of the mask grid and goes through the plate off-normal. The OPEN position has a diameter of 20 mm, so the unmasked beam is defined instead by the 19-mm-diameter racetrack baffle except on the elongated side to the right as viewed from the front.

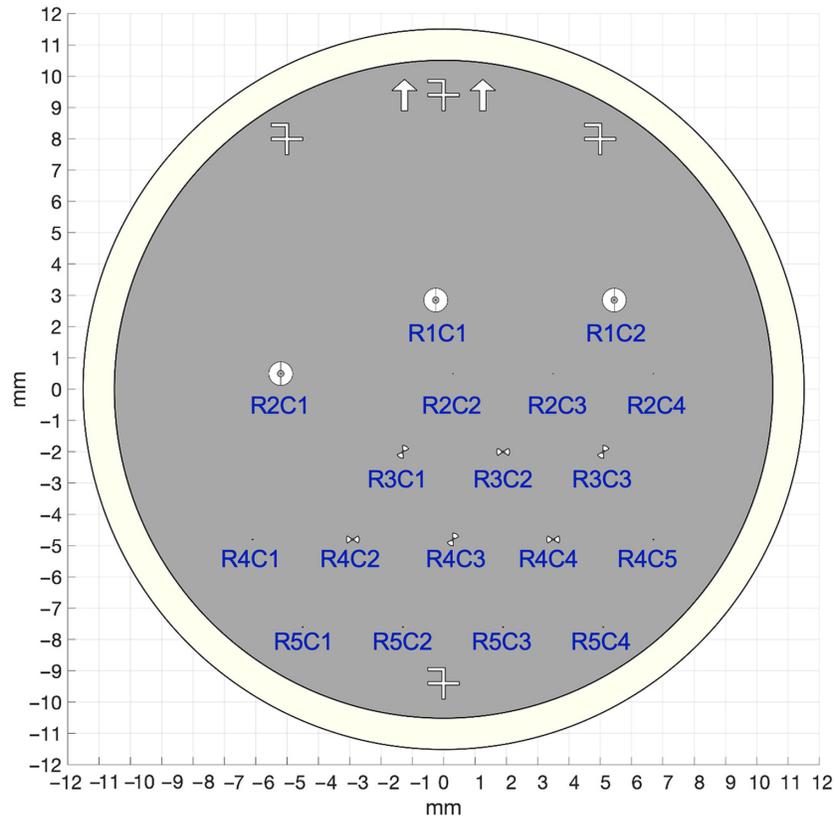

**Fig. 12** Transmissive SPC12 substrate at FPAM with positions labeled as used in Tables 1 and 2. The gray areas are coated with aluminum, and the outer 1 mm in radius is unpolished and uncoated. This substrate carries the spectroscopy and rotated spectroscopy FPMs for SPCs in band 1, the wide FOV FPMs for the SPC in band 1, and calibration pinholes for band 1.

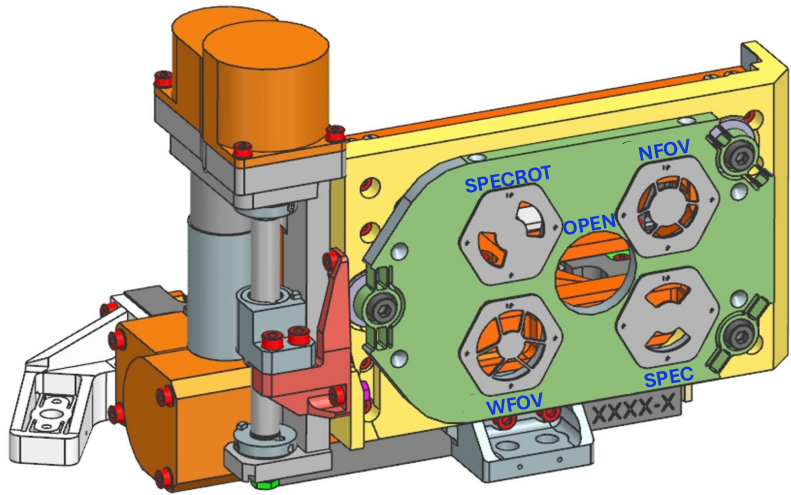

**Fig. 13** Lyot stop alignment mechanism (LSAM) in the Roman Coronagraph Instrument. The five nominal positions are labeled. The four Lyot stops are fabricated as separate devices, and the open position is a hole in the underlying optical interface plate.

As can be easily seen by the shapes of the Lyot stop openings, the numerical optimizations of each coronagraph mode favor very different Lyot stop geometries. The spectroscopy SPCs apodize and spatially filter the beam so much that the struts are no longer part of the Lyot plane illumination at all and nearly all the remaining off-axis light in the Lyot-plane passes through the bowtie-shaped openings. The wide FOV and multi-star modes also spatially filter the beam such that struts disappear in the on-axis PSF at the Lyot plane, but the struts were kept (though not oversized) in the Lyot stop because the central obscuration still needs mechanical support in the through-hole mask. The HLC Lyot stop is the only one following the traditional design methodology of having all the telescope obscurations oversized.

For someone choosing which Lyot stop to use with the unsupported, conventional (i.e., low contrast) Lyot coronagraphs, the two options are the Lyot stops having struts. The wide FOV Lyot stop offers higher throughput because of obscurations that are less oversized with the annular part of the opening extending from 36% to 91% of the beam diameter. The downside is that the smaller obscurations, including the struts not being oversized at all, might hurt the contrast significantly. The narrow FOV Lyot stop has a narrower opening from 50% to 80% of the beam diameter and pads the struts by 0.2% of the beam diameter on each side.

As for where each of the masks should be placed in the five available positions, there was a lot of flexibility. The two required positions are the ones named OPEN and NFOV, so one of those should go in the center where there is the most travel buffer. Having the open position in the center is probably better mechanically to keep it away from the flexures. The four corner positions all have similar travel tolerances of a few hundred microns past each mask center, which is relatively large, so any arrangement of the masks was equally valid.

### 7.3 Fiducials

Because light is diffracted outside the nominal beam diameter at the Lyot plane, the up arrow and "*F*" fiducials could not be through-holes in the substrate. Instead, they were made as shallow etches in the 10-micron-thick top layer, called the device layer, of the silicon-on-insulator (SOI) wafers used. These are visible by eye and under a microscope but have no effect on the transmitted beam.

### 7.4 Late Re-design to Include Fillets

In the final month before the mask designs were finalized, a mechanical issue was found in the design of the Lyot stops and field stops. Both types of masks are made as through-holes in SOI wafers, with an overhang of the 10-micron-thick device layer defining the optical edge. In the testbeds, we had routinely made these parts with sharp corners. For flight, however, we had not accounted for launch vibrations causing cracks at these high-stress points.

All four Lyot stop designs were numerically re-optimized to include rounded corners. The main trade was fillet radius versus core throughput of the PSF. Fortunately, there was a clear break point for each design that allowed for sufficient corner rounding and negligible loss in core throughput. A fillet radius of 2% of beam diameter (340 microns) was selected for the NFOV and WFOV Lyot stops and 3% (510 microns) for the SPEC and SPECROT Lyot stops. The field stops are much smaller openings—some only $1\lambda/D$ wide, so the maximum possible fillet radius was only 30 microns in all but one case.

During TVAC, which occurred after vibe tests, all Lyot stops except for SPECROT and several field stops were imaged on EXCAM. No damage such as cracking was found, so this risk appears to have been retired.

## 8 FSAM Masks

The final mask-carrying PAM is FSAM. As shown in Fig. 14, FSAM has two through holes for open positions, a backup ND filter with an optical density of 4.75, and a silicon wafer containing 31 field stops and a dark position. The FSAM ND filter is only for rare situations when an FPM is in use and the FPAM ND filters cannot be used; the change in path length from the substrate defocuses the beam compared with all the other FSAM positions, which are through holes. Because of limited space in front of FSAM, the PAM was installed facing downstream with the mask array bonded face down and the fixed baffle installed 14.5-mm downstream. The baffle being installed downstream of the openings is not ideal for stray light but was the only option.

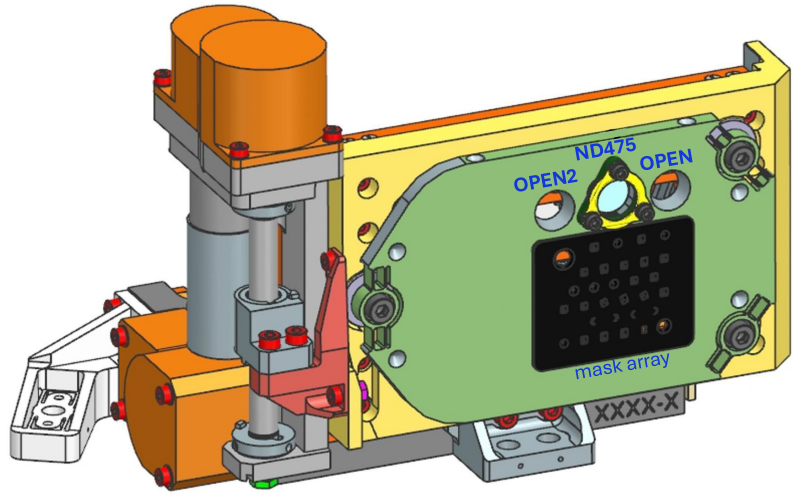

**Fig. 14** Field stop alignment mechanism (FSAM) in the Roman Coronagraph Instrument. The three upper positions and the silicon wafer containing all the field stops are labeled. Unlike the other PAMs, the field stop mask array is mounted face down on FSAM, and the baffle is after the PAM due to space constraints on the bench.

## 8.1 FSAM Layout Design Constraints

The final FSAM mask array as viewed from upstream is diagrammed in Fig. 15 with all the positions labeled. As with the Lyot stops, the entire device was etched out of an SOI wafer. The outer dimensions are 36.00-mm wide by 29.00-mm high. Because of travel restrictions from the PAM stages, needing a minimum of 2 mm around the outer border for handling and bonding the device face down, and needing 1 mm of additional buffer to accommodate up to 2 arcsec of on-orbit boresight alignment, the usable area of the device was reduced to an inner rectangle 23.00-mm wide and 21.14-mm high.

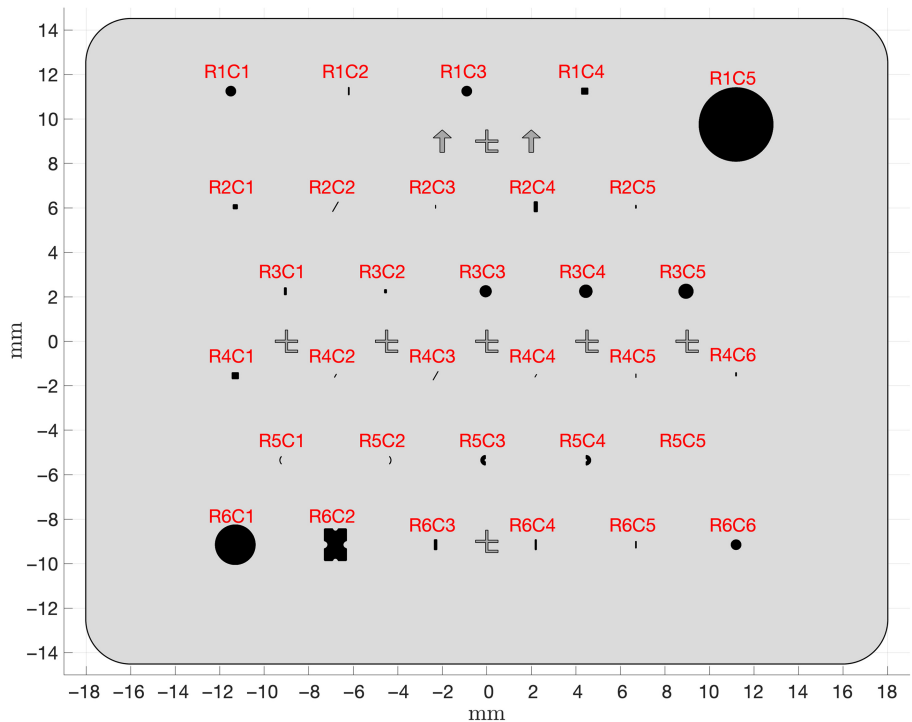

**Fig. 15** The silicon wafer containing all the field stops in the instrument. The "*F*" and arrow fiducials are shallow etches on the front side, not through holes.

The FSAM baffle was chosen to have the same field of view as the FPAM baffle. The different plate scale of 2.1 arcsec/mm at FSAM (versus 2.7 arcsec/mm at FPAM) makes the FSAM baffle radius larger at 4.33 mm. The FSAM baffle is a simple circular opening. Ideally, all neighboring field stops would be blocked by the baffle when one is in use, but doing so would limit the number of field stops. To fit in all the desired field stops, we opted only to use the conservative mask spacing of ≥4.90 mm, which includes tolerances, in the first row. For the best-effort, unsupported, or spare masks in the other rows, we opted for the smaller inter-mask separation of ≥3.8 mm corresponding to the 8.0 arcsec FOV of the DPAM baffle downstream of FSAM.

### 8.2 Specifics of Each Field Stop

Because the field stops are not diffractive and are re-imaged directly onto EXCAM, there was a lot of flexibility in their shapes and sizes. The final set of 31 field stops as listed in Table 3 was iterated and agreed upon by a multi-institutional working group of scientists and engineers in 2020. Only three positions are exact spares of others to ensure key capabilities. The rest have unique shapes and sizes, but most are similar to at least one other to serve as a backup if needed. All of the field stops fit into these five categories:

1. Small circles for narrow FOV imaging with the HLCs.
2. Large circles for polarized imaging.
3. Slits for spectroscopy.
4. Squares for multi-star imaging.
5. Half annuli for one-sided imaging in case of a single DM failure.

First, we will describe some of the arrangement decisions before discussing the mask properties. The two large circular openings were placed at opposite corners of the device because they needed larger buffers than the small openings. The bottom row is near the lower vertical travel limit of the stage, so only lower importance masks were placed there in the unlikely event that misalignments were to make it unreachable. There was some concern about what to do if a single PAM stage were to stop functioning, so some key masks (at least one HLC band 1 field stop, a slit, and the largest opening) were put in both row 1 and in the rightmost column. However, this strategy partially failed because the large opening at R1C5 ultimately had to be moved down and out of line with the rest of row 1 to have enough of a border at the edge of the device. As stated before, the important masks in row 1 had a larger spacing than for the other rows. The other two multi-star field stops not in row 1 were placed along the left edge to help reduce the chances of stray light from the off-axis star getting through. Otherwise, the rest of the field stops are all relatively small and were placed at will in the remaining positions.

The six circular openings with diameters ≤1400 mas are intended for narrow FOV imaging with the HLCs. The HLC field stop sizes are chosen as $9.7\lambda/D$ in radius to account for the TTR5-required $9.0\lambda/D$ outer working angle plus several tolerances totaling $0.7\lambda/D$. The tolerances are as follows, with $\lambda$ indicating wavelength-independent effects and $\lambda_1$ indicating effects calculated for band 1. There is sometimes a 1-pixel-wide arc of pixels along one edge of the field stop that is lost to glint, equivalent to $0.43\lambda_1/D$. To account for a 1% uncertainty in the beam diameter from the observatory, we add another $0.10\lambda_1/D$. Finally, the requirement on star position of 1/6 a pixel and the PAM step size of ≈1/5 a pixel add another $0.16\lambda_1/D$. The unsupported HLCs for bands 2, 3, and 4 have no requirements, but they were provided with the same $9.7\lambda/D$ radii field stops for consistency.

At the corner positions, R1C5 and R6C1 are the two large circular field stops for polarized imaging. The Wollaston prism modules split the beam with centers separated by 7.5 arcsec. The 3.5-arcsec radius circle at R1C5 is meant to be the baseline field stop for polarized low-contrast imaging to avoid crosstalk between the two beams and include some margin. It also captures almost the entire unvignetted FOV in regular imaging, which could be used if stray light is found to be an issue at the OPEN position during in-orbit calibrations. The backup position for polarized imaging at R6C1 has only a 1.9-arcsec radius to match the unvignetted FOV of the polarization modules in case stray light is found to be a problem for those.

There are 17 slits of various shapes, sizes, and orientations to cover many possible use cases. The major considerations for most of the slit widths were the PSF width (either the FWHM or the distance between the first nulls on opposite sides of the star) and whether it could work with the contributed HLCs.

A major functional difference of using one of the contributed HLCs instead of an SPC for spectroscopy is that the HLC needs a field stop to block the incredibly bright (up to $10^{-2}$ contrast) speckles outside of the $9.7\,\lambda/D$ radius dark hole from being seen on EXCAM. When the nominal, circular HLC field stop is replaced with a slit, that slit must also not allow in any speckles from outside the original field stop opening. For this reason, there are many vertically "short" slits of $6\,\lambda/D$ to be compatible with the HLC. There are only two dedicated rectangular HLC slits with widths sized at $1.1\,\lambda/D$ for the HLC PSF's FWHM—one for band 2 in position R2C3 and one for band 3 in position R4C5. The rest of the rectangular, vertical slits compatible with the HLC are so-called "short" versions of the wider SPC slits; these also serve as backups for the "tall," nominal use case SPC slits. In case none of the other slits work in a given situation for the HLC, there are also two unconventional quarter-annulus slits at positions R5C1 and R5C2 to allow coverage to the very left and right edges, respectively, of the HLC dark hole.

The SPCs are expected to provide better performance for spectroscopy because of lower sensitivities to low-order aberrations, so most of the slits were designed for the two orientations of the spectroscopy SPCs. As shown in Fig. 16, these heavily apodized SPCs have a nonstandard PSF shape. The main lobe is elongated by about a factor of two compared to the unmasked PSF (i.e., the FWHM is $\approx 2\,\lambda/D$ and the distance between first nulls is $\approx 4\,\lambda/D$) along the same axis as the bowtie-shaped dark hole. Perpendicular to that axis, there are two bright secondary lobes with peaks over half as bright as the main lobe. For the nominal orientation spectroscopy SPC, only the vertically oriented slits are used to disperse light horizontally from all three of the companion's PSF lobes. For the rotated spectroscopy SPC, only the rotated slits are used to isolate light from the just the central PSF lobe.

The orientations of the slits for nominal and rotated orientations are not interchangeable. In either case, the slits were designed to avoid any crosstalk among the bright SPC PSF lobes. The two orientations of the spectroscopy SPCs are included to provide approximately twice the available on-sky coverage on any given date, not to provide alternate choices for which PSF-slit combination to use. In a chosen observation window, it is expected that an exoplanet would only be within the field of view of one of the two orientations of the spectroscopy SPCs.

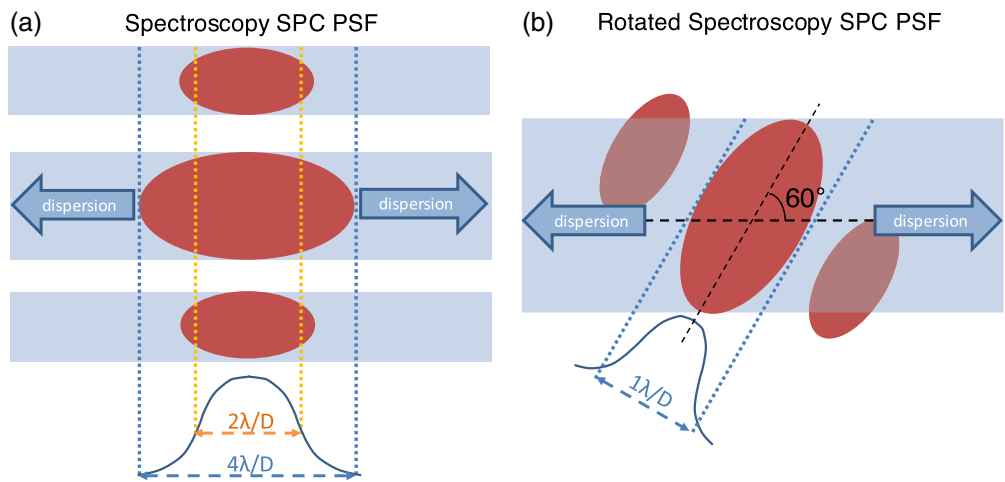

**Fig. 16** Diagram illustrating how the slit widths and orientations were chosen with respect to the nonstandard SPC PSF, both in the best-effort nominal (a) and unsupported rotated (b) orientations. Due to the heavy apodization at SPAM, the spectroscopy SPCs have an elongated primary lobe and two secondary lobes with peaks over half as bright as the main lobe. For the nominal orientation SPC, two slit widths were included: one to capture the entire main lobe ($\approx 4\,\lambda/D$) and another for just the FWHM ($\approx 2\,\lambda/D$) of the main lobe. The axis of the three lobes is perpendicular to the prism dispersion axis, so light from all the lobes can be used. For the rotated SPC, it was determined that sampling just the main lobe along its narrow axis ($\approx 1\,\lambda/D$ wide) with a rotated slit would be the most efficient and avoid crosstalk among the three lobes of the PSF.

The slits intended for the most usage are the ones sized for the FWHM of the PSF ($2\lambda/D$) in the nominal orientation and have heights of $12\lambda/D$ to cover the entire bowtie dark hole even at its largest span. The field stop for band 2 is in position R6C5, and the one for band 3 is in position R1C2. The short $6\lambda/D$ high versions of these for double use with the HLC are in positions R2C5 for band 2 and R4C6 for band 3. The null-to-null $4\lambda/D$ wide slits are lower priority and are included in case misalignments or drifts are larger than expected or to get more exoplanet throughput if the stellar speckle background is relatively low. There is just one tall version of this field stop at position R3C1 and one short one at R3C2, so they are sized for a compromise wavelength (695 nm) halfway between the centers of bands 2 and 3.

The rotated spectroscopy PSF cannot be sampled the same way because the dispersion axis cannot change. Collecting light from all three lobes would result in crosstalk, so it was decided to isolate only the main lobe of the PSF with a rotated slit. The width for all of these was chosen as $1\lambda/D$ to span first nulls and still be wide enough to use with the HLC PSF. There are long and short versions of these slits. The long one is $17\lambda/D$ to mostly span both sides of the SPC dark hole but also fit inside the HLC dark hole at more positions, and the short one is again $6\lambda/D$ long.

The mask designs had to be finalized in 2020 before Astro2020 could advise on whether a starshade rendezvous with the Roman Space Telescope should be prioritized (it was not), so two additional slits were included for that possibility. A starshade would not require any masks within the instrument to be used, so the main consideration for the slits was their width to accommodate the maximum linear displacement of several known habitable zone exoplanets over an expected 25-day observation window. It was determined that slit widths ranging from 120 to 300 mas in increments of 60 mas suited nearly all of the chosen targets. This plan only required two new slits and reused the existing 127-mas-wide slit at R1C2 and 243-mas-wide slit at R3C1. The two starshade slits added are the 180-mas-wide one at position R6C4 and the 300-mas-wide one at position R6C3.

There are three rounded square field stops for multi-star imaging should such a HOWFSC algorithm be developed for the Roman Coronagraph Instrument. The ones at R1C4 and its spare at R4C1 are sized for a $9 \times 9\, \lambda_4/D$ dark hole as chosen by the multi-star designers.[31] The multi-star SPC also has a band 1 option, so there is also a single $9 \times 9\, \lambda_1/D$ field stop at position R2C1. In retrospect, it may have been better to use the spare position at R4C1 to include an even smaller field stop option (e.g., $5 \times 5\, \lambda_1/D$) to make MSWC easier as has been favored in recent vacuum experiments.[32]

In the unlikely, catastrophic event of a single DM failure, three field stops were included to enable one-sided dark holes (i.e., over only 180 deg of the FOV). Without blocking the uncorrectable side of the image, the dynamic range would otherwise be too large for the detector. Positions R5C3 and R5C4 are the left and right halves, respectively, of an annulus open from 2.3 to $9.7\, \lambda_1/D$ to be used with the required HLC in band 1. Position R6C2, which looks like a piece of a jigsaw puzzle, is the union of half annuli for the wide FOV SPC in bands 1 and 4. This combined field stop is possible because the SPC FPMs already have a built-in outer field stop, so there is no outer blocking needed at FSAM. The top and bottom of the jigsaw puzzle piece are sized for band 1, and the left and right are for band 4. The annular radii are $1\lambda/D$ more open than the FPMs to allow easier alignment and extend from 4.6 to $21.4\, \lambda/D$ in bands 1 or 4. For single-side dark holes with the spectroscopy and rotated spectroscopy SPCs, additional field stops were not needed because the $9 \times 9\, \lambda_4/D$ MSWC squares are just the right size to be re-used for this purpose.

## 9 Flight Mask Devices

In this section, we show photographs of all the flight mask devices fabricated from 2020 to 2022 and selected for the Roman Coronagraph Instrument. For each device, we describe some of the key measurements showing that all mask fabrication requirements needed to meet the $\approx 10^{-9}$ contrast performance goals were met for the required and contributed mask configurations. Some measurements are also reported for unsupported masks, for which there were no requirements. All of the flight masks were fabricated in JPL's MDL. JAXA provided the custom mask substrates needed for SPAM and FPAM, and the SOI wafers for LSAM and FSAM masks were commercial, off-the-shelf parts. All of the masks were fabricated within their necessary tolerances, sometimes much better.

### 9.1 SPAM Mask Array

The mask array for SPAM was the most difficult to fabricate with over 40 steps required. There were only two candidate flight devices, with one exhibiting micron-sized spots scattered across all the aluminum regions. The selected flight device was free of those spots and met all its low-level requirements. The fine features of the flight SPM array were verified using high-resolution, stitched microscope images such as the one in Fig. 17(a). The device is shown mounted on SPAM in Fig. 17(b).

The 4-mm-thick silicon substrates provided by JAXA were superpolished and the wide temperature ranges during SPM fabrication did not alter the surface substantially. Figure 18 shows the Zygo-measured surface figure of the flight device with the tip and tilt removed. Table 4 lists the relevant Zernike polynomial decomposition measured while zoomed in on each SPM position. The surface figure errors for each mask are several times less than the required values of <5 nm RMS for defocus and <25 nm RMS for all terms above defocus. These small surface figure errors are important because the small openings in the SPMs cannot be resolved well enough for wavefront flattening of high-order terms.

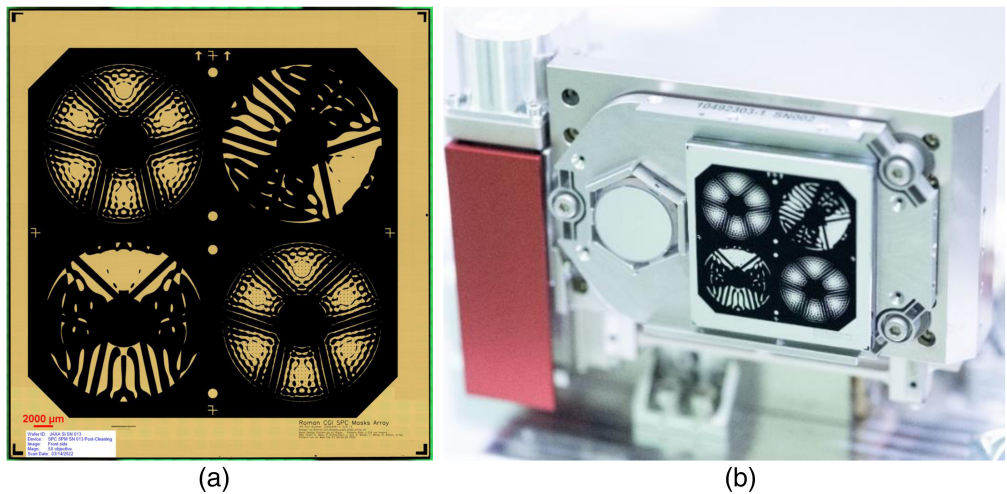

**Fig. 17** Manufactured flight mask device for SPAM (a) as imaged under a microscope and (b) mounted on SPAM.

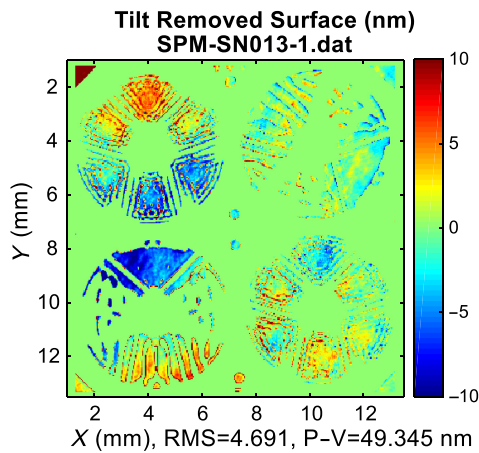

**Fig. 18** Surface figure measurement of the SPM mask array before mounting to the SPAM OIP. Tip and tilt terms have been removed.

**Table 4** Surface figure error (SFE) for each of the SPMs. All four SPMs are well below the requirements of <5 nm RMS SFE of defocus and <25 nm RMS SFE for everything excluding piston, tip, tilt, and defocus.

|                  | WFOV | SPEC | SPECROT | MSWC |
|------------------|------|------|---------|------|
| Z4 RMS SFE (nm)  | 0.1  | 0.9  | 0.8     | 0.3  |
| Z5+ RMS SFE (nm) | 3.9  | 3.0  | 3.5     | 4.0  |

The black silicon and aluminum both met their necessary reflectance values. The bare aluminum was required to have >83% reflectance across all the instrument bandpasses. A witness sample fabricated at the same time as the mask array was measured to have ≥84.8% reflectance with a PerkinElmer Lambda 1050 spectrophotometer. The hemispherical reflectance of the black silicon was measured to be <0.3% over the wavelength range 600 to 900 nm using the same spectrophotometer; the requirement was <0.5%. The specular reflectance of the black silicon was measured as $3.6 \times 10^{-8}$ at 633 nm, which meets the required $<1.0 \times 10^{-7}$ and is expected to contribute incoherent light in the dark hole at a miniscule $9 \times 10^{-11}$ contrast for the spectroscopy SPC and $4 \times 10^{-12}$ for the wide FOV SPC. The experimental setup for specular reflectance measurement and method of calculating the expected incoherent contribution to the dark hole is the same as that described in the milestone 1 report for the project.[39]

The edge tolerance of the 17.0-micron-wide pixels of the SPM designs was measured as <0.1 microns using several different feature measurements at high magnification under a microscope. The requirement was much looser at ±1.0 microns. For reference, the WFOV, SPEC, and SPECROT SPMs have 1000 × 1000 pixel grids; the MSWC SPM has a slightly smaller 982 × 982 pixel grid to obtain the desired spacing for the integrated dot pattern.

The entire SPM array was imaged at high magnification under a microscope to check for defects. Finding all the defects in the images by eye is time-consuming and error-prone, so we developed an interactive MATLAB script to detect SPM defects. First, the user clicks on four equivalent points in the stitched microscope image and an ideal representation of the SPM. Then, the script uses tools in MATLAB's Computer Vision Toolbox to co-align the measured image with the design and then detect any differences. All the potential defects are highlighted in the measured image, and the user has to approve or disapprove each one. Finally, the total area of the approved defects is computed. The total defect areas computed in each SPM for the flight device are 1597 $\mu m^2$ in six defects at the WFOV position, 1518 $\mu m^2$ in three defects at SPEC, 3085 $\mu m^2$ in four defects at SPECROT, and 385 $\mu m^2$ in one defect at MSWC. The largest defect on each SPM is shown in Fig. 19. Most of the defects are blobs of black silicon or flakes of aluminum—presumably from the post-etching wafer dicing process—that have settled elsewhere as is shown for the WFOV and SPEC positions. The largest defect is on the SPECROT position and appears to be where a pool of debris dried out at the end of a liquid cleaning.

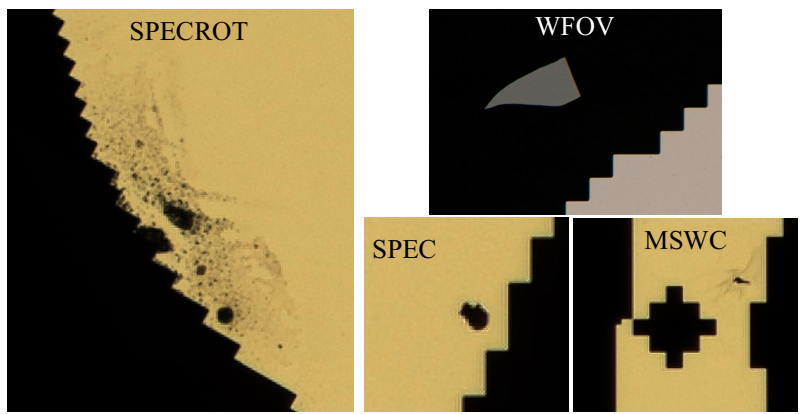

**Fig. 19** Zoomed-in microscope images of the largest defect on each SPM. Not to scale. For reference, the smallest step size in the pixellated SPM pattern in each image is 17 microns.

The contrast degradation from these defects should be small due to their small sizes diffracting light to large angles and their total area covering $\approx 10^{-5}$ of the pupil area. Assuming a particle size of $\approx 1/1000$ the pupil beam diameter (based on Fig. 19), thereby giving a PSF core area of $\approx 10^6 (\lambda/D)^2$ at $\approx 10^{-5}$ the total brightness of the nominal beam, the estimated average contrast over the dark hole would be small at $10^{-11}$. To be certain and there were only a small number of defects, we manually constructed representations of the best-effort SPMs with these defects and computed the open loop contrast degradation. We only considered the amplitude and not the phase of the defects as that was not measured. To include each defect in the apodizer model, we cropped and downsampled the measured image around the defect to match the position and scale of the designed SPM, normalized both images, and subtracted. We then further isolated the defect to avoid slight mismatches along the designed feature edges and added the defect to the designed apodizer. Running these modified SPMs through a simple open-loop model of the coronagraphs resulted in small contrast degradations close to the rough estimate: $6 \times 10^{-11}$ for the spectroscopy SPC and $2 \times 10^{-11}$ for the wide FOV SPC.

### 9.2 FPAM Mask Arrays

The FPM mask devices had to be fabricated well for both the HOWFSC beam path in transmission and the LOWFSC path in reflection. In addition, the mask layers must be co-aligned well enough (the requirement was ±1 micron for the band 1 HLC and ±2 microns for the SPCs) for both LOWFSC and HOWFSC to work at the same time. The centering of the mask layouts on the substrates must be ≪1 mm not to erode the on-orbit boresight alignment budget, and the measured centerings were found to be <0.1 mm for each substrate. Stitched microscope images of the final four FPAM mask substrates are shown in Fig. 20. The mask substrates and the two FPAM ND filters are shown bonded to the FPAM OIP in Fig. 21. The only significant differences between the mask substrate designs and manufactured layouts are (1) metal pads on the left and right edges added to offload charge build up during the lithography steps and (2) text blocks along the outer edge containing identifying information. The metal pads and text blocks are outside the FOV of all the coronagraphic masks.

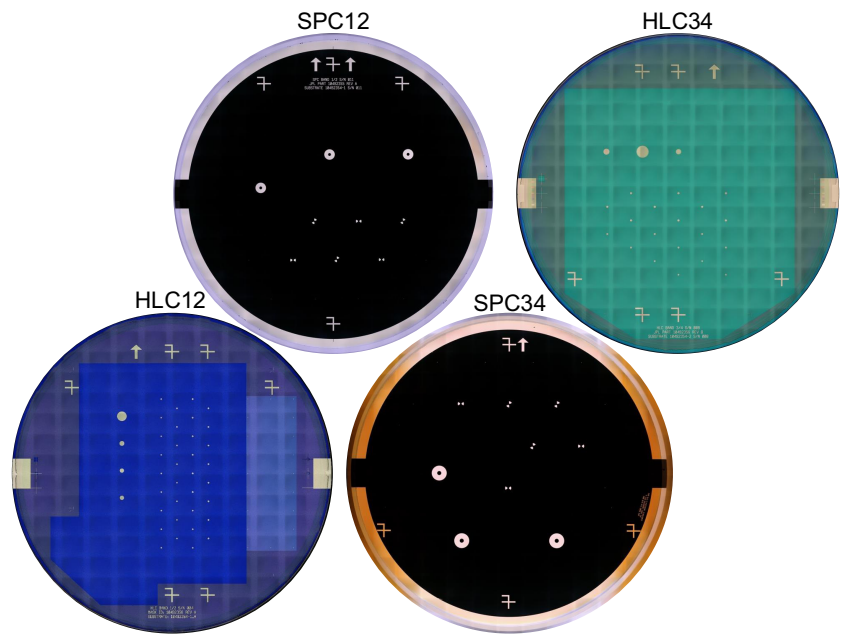

**Fig. 20** Stitched microscope images of the four FPAM mask devices selected for flight.

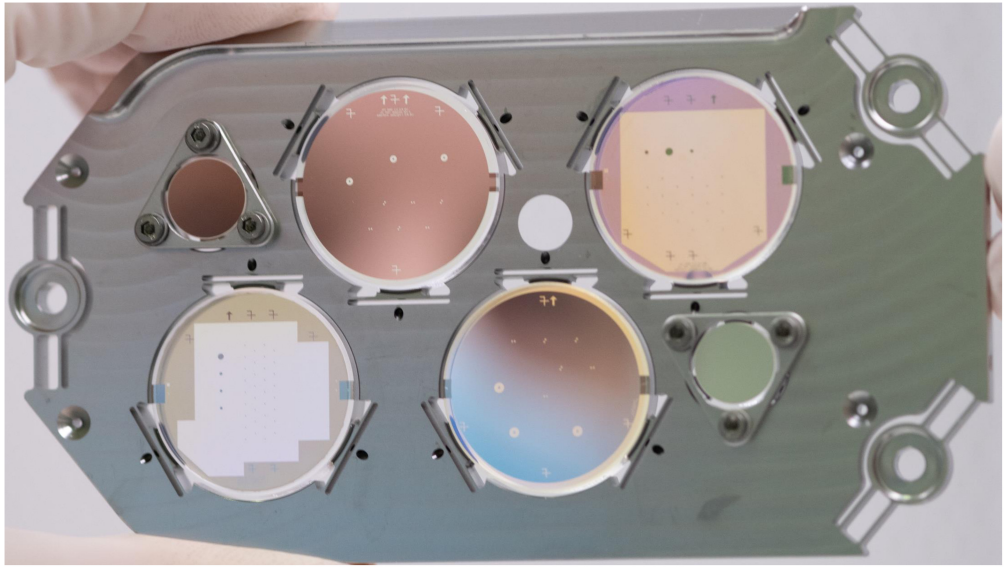

**Fig. 21** Photo of the four FPAM flight mask substrates and two ND filters after being bonded to the FPAM optical interface plate.

### 9.2.1 *HLC12 substrate*

The most important of the substrates and first fabricated was HLC12, which carries the required band 1 HLC narrow FOV occulters. These band 1 occulters have the most complicated PMGI pattern and had to be made extremely accurately for the HOWFSC and LOWFSC performance requirements to be met. The C2R5 occulter on the HLC12 substrate, measured with an atomic force microscope (AFM) as shown in Fig. 22, has near-perfect PMGI morphology with $+25$ nm of bias error and no peak-to-valley error. This occulter was the one used to meet the HOWFSC and LOWFSC requirements in TVAC. The co-registration of the PMGI layer and nickel disk underneath was measured to be better than 0.1 microns using Vernier scales along the edges of the substrates—much better than the $\pm 1$ micron tolerance needed for the desired HLC HOWFSC performance with the LOWFSC pointing control loop locked and centered on the PMGI pattern. The underlying metal disk of the occulter is designed to be 3 nm of titanium as an adhesive layer with 109 nm of nickel on top for 112 nm total. It was measured to be 114 nm in total, which results in negligible performance degradation for HOWFSC and LOWFSC.

The unsupported, dual-path ZWFS spots in column 8 of the HLC12 substrate are designed to work simultaneously for HOWFSC and LOWFSC on a bright star, with LOWFSC using several percent of the starlight reflected from the PMGI. The dual-path ZWFS spots were accidentally designed without accounting for the substrate's AR coating. In that simpler scenario, the thin-film effects are negligible. Any bias thickness of the PMGI would work, and the phase-shifting dimple should be $216 \pm 20$-nm deep and $1.0\,\lambda/D$ (18.5 microns) in diameter. The fabricated device has AFM measurements of 1530 nm for the PMGI bias at the edge of the substrate and dimple depths in rows 2 to 7 of 174, 190, 204, 214, 231, 244, and 241 nm. The different depths were intentional to allow for possible fabrication tolerances and modeling errors. Newer analysis using an approximation of the proprietary AR-coating shows that the ZWFS response—both in reflection and transmission—is much more nonlinear chromatically than the originally assumed model of PMGI directly on fused silica. The as-manufactured, dual-path ZWFS will be less responsive at some wavelengths in the relatively large, 128-nm-wide LOBE bandwidth but will probably still work overall for differential phase measurements.

On the large-diameter Lyot occulters, the desired dimple depth for LOWFSC is different because of the underlying nickel disk. The target was 130 nm and the AFM-measured value was 134 nm, and the PMGI bias thickness was allowed to remain at the original, pre-etching value.

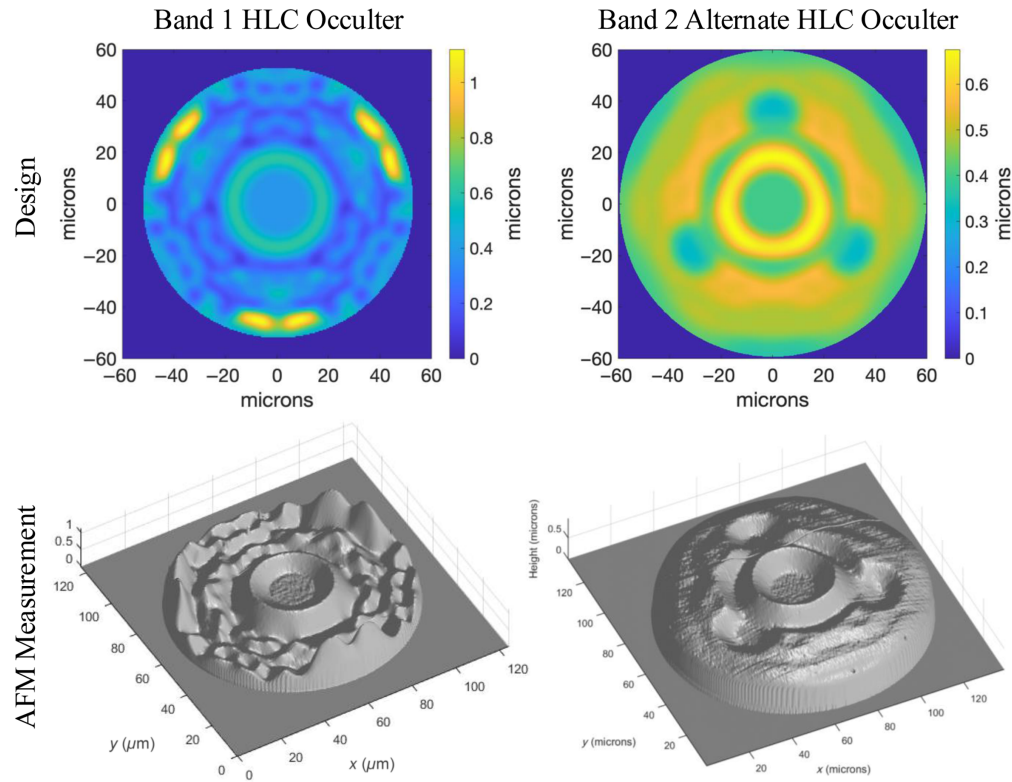

**Fig. 22** Designed (top row) and AFM-measured (bottom row) occulter profiles of the required band 1 occulter HLC occulter HLC12_C2R5 (left) and the unsupported, alternate band 2 (right) HLC occulter HLC12_C4R5. The band 1 occulter's dielectric layer was patterned extremely accurately with 0 peak-to-valley error and only a 25-nm bias error.

#### 9.2.2 HLC34 substrate

All of the masks on the HLC34 substrate are unsupported but were still fabricated quite accurately. The complicated occulter profiles for the band 3 and alternate band 4 HLCs are shown in Fig. 23. The deviations show a steep droop at the outer edge and a small positive error over the interior of the profile. Because the original HLC occulter designs were optimized for only HOWFSC before the LOWFSC capture range requirement was included in the mask design evaluation, it was by chance that the bands 1 and 3 occulter designs worked with LOWFSC without any changes. The now-alternate band 2 and 4 occulters, however, did not because outside the ridge surrounding the starting dimple, there is a trench of a similar height to the dimple. If the star starts inside the ridge at a radius of less than 45 mas ($\approx$16 microns), LOWFSC can center the star, but if the star starts outside the ridge it cannot.

The simple profile HLC occulters for bands 2 and 4 were designed to have an inverted top-hat PMGI profile typical of a ZWFS. Because of thin film effects, the PMGI bias thickness had to be chosen carefully to work well for both LOWFSC and HOWFSC. LOWFSC responsivity had local maximums for PMGI biases of 350, 540, 740, 940, and 1160 nm with a tolerance of $\pm 50$ nm at each and assuming a dimple depth of 130 nm. The FALCO[40] software package was then used to perform HOWFSC for occulters with PMGI thicknesses finely sampled across these five windows. The PMGI bias value that provided the best balance of contrast, DM stroke usage, core throughput, and jitter sensitivity for the HLC designs was found to be 540 nm independently for both bands 2 and 4. The low-contrast occulters do not use HOWFSC but do use LOWFSC, so they were designed to have the same PMGI thickness as well. Using an AFM, one of the band 4 occulters was measured to have errors of just $+16$ nm for the PMGI bias thickness and $+8$ nm for the dimple depth. The underlying titanium and nickel layers were designed to be 3-nm and 100-nm thick, and the total metal height measured with an AFM was 106 nm.

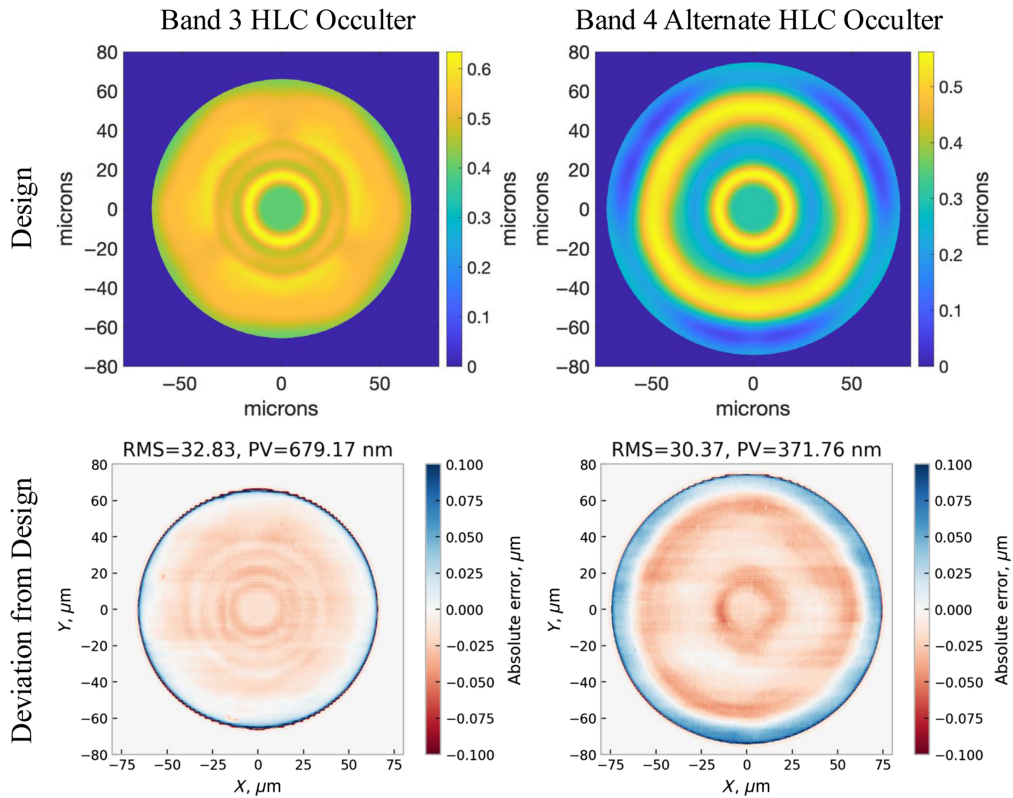

**Fig. 23** Designed occulter profiles (top row) and AFM-measured deviations (bottom row) occulter profiles of the unsupported band 3 HLC occulter (left) and the unsupported, alternate band 4 HLC occulter (right).

### 9.2.3 *SPC12 and SPC34 substrates*

The SPC FPM devices were simpler to fabricate as they required no grayscale etching, just two uniform-height layers of aluminum. Key parameters were the FPM lateral dimensions, opacity in the aluminized areas, LOWFSC spot height, LOWFSC spot lateral size, and FPM-to-LOWFSC-spot coalignment. The lateral dimensions and coalignments were measured to have accuracies of 1 micron or better using manually fit circles and ellipses on 500× magnification images with an uncertainty of about 1 pixel or 0.43 microns. The LOWFSC spots were additive rather than recessed so that the transmitted light would be suppressed more than in surrounding areas. The target height was 72 nm and was measured at 70 nm on the SPC34 substrate using an AFM. The goal opacity for the aluminized areas was an optical density of at least 6, and the actual value is ∼11 for the measured thickness of 3 nm of titanium with 165 nm of aluminum on top.

### 9.3 LSAM and FSAM Masks

The Lyot stops and field stops mask array are all freestanding, through-hole masks. To achieve the desired lateral tolerances for the optical edges of ≪0.1% beam diameter (≪17 microns) for the Lyot stops and <2 microns for the field stops, we had to use microlithography to fabricate both. We were able to fit two copies of each Lyot stop and six copies of the field stop array on each of the three SOI wafers processed as a whole. One of them is shown in Fig. 24 after depositing the pattern but before etching. The four masks and open position at LSAM could have been fabricated as a single mask array, but it was decided to keep them separate to improve the fabrication yield.

The optical edges of the masks are defined by the 10-micron-thick device layer of the SOI wafer. In the testbeds, we have used device layers as thin as 1 micron to reduce edge glint, but to survive the stresses of launch vibrations, we opted for the thicker 10 microns. The recessed backbone of the masks was patterned with oversized holes etched into the 640-micron-thick handle layer of the wafer.

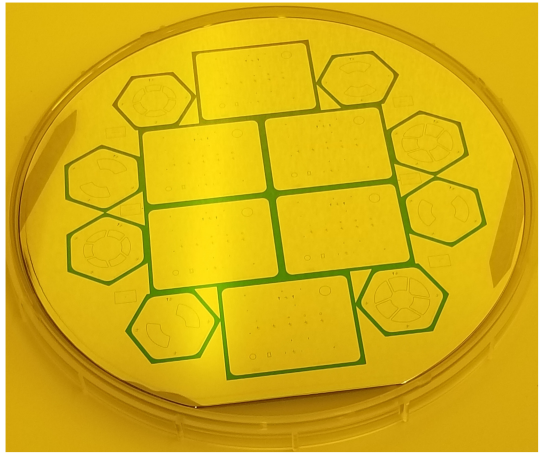

**Fig. 24** Silicon-on-insulator wafer patterned via microlithography and later etched to produce two instances of each Lyot stop and six of the field stop array.

The four selected Lyot stops for flight are shown in Fig. 25. In these stitched microscope images, green is transmitted light, yellow is light reflected off the aluminum coating, and brown is light reflected off the uncoated silicon. The uncoated areas at the edge of each device are where it was either taped or clamped down in the coating chamber. The uncoated regions also provide a place to pick up the devices with tweezers without damaging the aluminum coating. Although the thick parts of the wafer are opaque enough in all of the instrument's bandpasses (silicon becomes transmissive in the infrared), we wanted an opaque metal coating for the 10-micron-thick overhangs that define the optical apertures. The 210 nm of aluminum deposited has an optical density of 11 or higher in all the instrument bandpasses.

The selected flight Lyot stops had no significant defects, only microscopic, cosmetic spots in a few places that do not affect the transmitted beam in any way. Some of the spare devices had small lines at some of the optical edges that may have been cracked, so they were ranked lower.

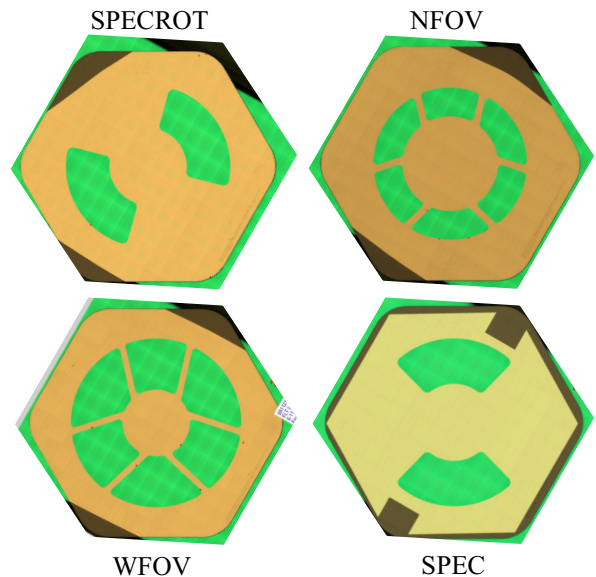

**Fig. 25** Stitched microscope images of all four Lyot stops chosen for flight. The openings are shown in green, the aluminized areas are yellow, and the uncoated silicon areas are brown. The rectangular pattern in the green area is an image stitching artifact. The uncoated regions along the outer edge are where the devices were taped or clamped down during coating and are needed as handling regions. The aperture dimensions matched the design values within measurement tolerance. The flight devices have no significant defects, only a few cosmetic spots.

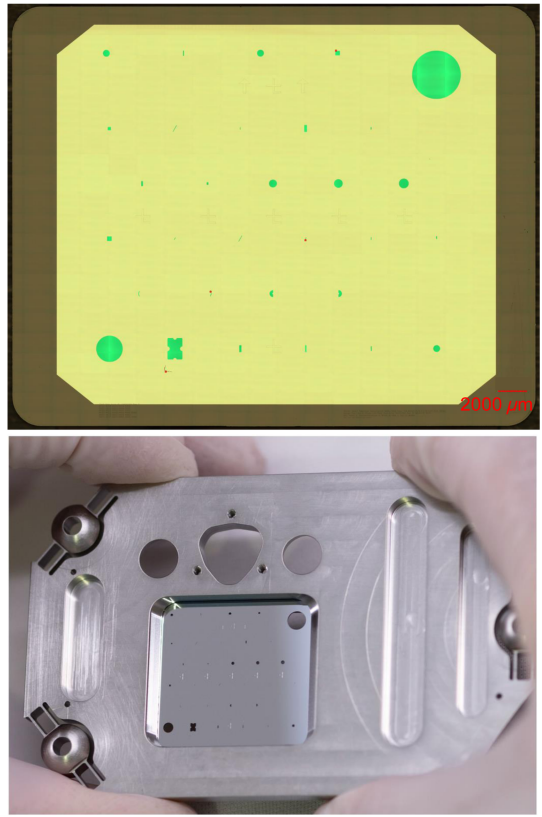

**Fig. 26** Stitched microscope images of the field stop array chosen for flight (top). A front-view photo of the same device after being bonded to the FSAM optical interface plate (OIP).

The hole sizes were difficult to measure precisely because the stitched images accumulate uncertainty at about 1 micron per frame. Because of this, the inner and outer diameters of the Lyot stops were found to be within the measurement tolerance ($\approx$10 microns) of the designed dimensions.

The field stop array selected for flight is shown in Fig. 26. The stitched microscope image is shown on the top. For this device, the entire outer perimeter was covered by a jig during coating so that the silicon could be directly bonded to the FSAM OIP as shown at the bottom of Fig. 26. This device had no defects for the required and best-effort masks and three negligible, micron-scale pieces of contamination along the edges of unsupported masks at positions R1C4, R4C4, and R5C2. There is also a piece of debris below the field stop at R6C2 that is not visible in transmission. The widths of 17 of the field stops were measured using high-resolution microscope images and found to deviate from the designed values by $0.6 \pm 0.1$ microns, much less than the required 2-micron tolerance.

## 10 Lessons Learned

One of the most important roles of the Roman Coronagraph Instrument is simply to serve as a trial run for HWO. Here, we describe some of the major lessons that we learned in transitioning from grant-funded testbed experiments in small teams to flight instrument development and testing as part of a large team. Some of these lessons have been mentioned in earlier sections, but we include them again here for easier reference. Although the numerical optimization of the masks is not detailed in this paper, we include the lessons learned while doing so for completeness.

### 10.1 Mask Optimization

1. The mask design team needs an automated pipeline that intakes a new coronagraph design and quickly returns a scientific yield. The most difficult part of optimizing the coronagraphic masks was a lack of a top-level metric that could be calculated quickly and fed back into the design process. The scientific yields from statistical yield calculators such as the Exoplanet Open-Source Imaging Mission Simulator (EXOSIMS)[41,42] or the Altruistic Yield Optimizer (AYO)[43] became too low to be statistically meaningful. Instead, we utilized a custom exposure time calculator using assumed properties of known exoplanets detected by the radial velocity (RV) method. At the time of the final SPC design in phase C, the exposure time calculator no longer had a scriptable interface. As a result, only three spectroscopy SPC point designs were able to be manually evaluated based on a time-to-SNR metric on fiducial exoplanets. This issue of integrated mask design and scientific yield calculation has largely been addressed for HWO with extensive use of the AYO package during early design development, but when requirements are eventually written, one should be made to guarantee this functional capability.

2. The repercussions of a major change to the coronagraph architecture take so long to realize throughout the instrument design and requirements that changes after Phase A are unrealistic. Another lesson we learned is that it takes at least a year for major design changes to propagate through the rest of the project. Not realizing that it was already too late, we continued researching avenues for large performance improvements late into Phase B. A radical change in the masks or DM configuration would require at least a year for design, fabrication, testing, and model validation in the testbed; updates to project requirements; and a partial re-design of the instrument. Realistically, the major coronagraph design choices for HWO will have to be made early in Phase A to be the baseline in the preliminary design review at the end of Phase B.

3. The observatory requirements must be defined with the coronagraph instrument in mind to avoid a moving target for the mask design and keep system-level performance out of mask design requirements. A mistake we made in defining the mask design requirements was setting coherent and incoherent contrast requirements at the subsystem level (Level 5) for mask designs instead of only at the instrument level (Level 4). For Roman, we chose to do this because we could not place any requirements on the observatory such as the pupil shape or the amount of jitter. The mask designs therefore had to adapt to absorb the system-level performance degradation from any changes to still meet the Level 5 system-level coherent and incoherent contrast requirements. The contrasts could only be computed using the latest—and therefore continually changing—closed-loop, end-to-end models of the instrument and observatory. Besides conflating system and subsystem level work, the other issue with this arrangement was that the mask design team had to maintain a parallel, system-level, rapid modeling capability independent of the dedicated integrated modeling team because the effort for that was not originally planned for. A better approach is to have Level 5 mask design requirements that depend only on information related to the masks (e.g., the ideal coherent contrast, no incoherent contrast except for polarization aberrations induced by the masks, sensitivities to mask misalignments, and sensitivities to low-order Zernike polynomials in the wavefront), leave the system-level modeling to the Integrated Modeling team, and have a project-wide agreement on how often the one official model is updated.

### 10.2 Designing Masks for the As-built System

1. Plan to make the coronagraphic masks after the observatory optics and other instrument optics have been fabricated, measured, and selected to account for the as-measured magnification and any defects. Because of schedule constraints and most hardware being developed in parallel, the mask designs had to be approved and fabricated before the rest of the optics were either characterized or selected. The two cases that mattered the most were the DMs and the observatory mirrors. We were extremely lucky that the dead DM actuators landed in tolerable locations behind the HLC Lyot stop and mostly behind SPM obscurations. Had they been fully in the unmasked parts of the pupil, the achievable contrast could have been heavily degraded. Depending on the telescope pupil and the coronagraph type, if dead or weak actuators are known before masks are designed, their impacts can be partially or sometimes even fully mitigated.

The later selection of the observatory mirrors presented another problem. All of the mirrors had been fabricated and measured, but the difference in beam magnification between the main set and the alternate set was 1%. This was large compared with the other terms in the magnification and translation error budget on the order of 0.1% each. Fortunately, this resulted in only minor throughput losses for the Roman coronagraphs mostly from the loss in area at the pupil outer edge. For a future mission such as HWO, any pupil mask designed to line up with the primary mirror segment gaps will want a magnification tolerance closer to 0.1% to maintain high throughput. One possibility for HWO might be to fabricate apodizers for several possible magnifications and then install at the last minute the one that works best for the finalized telescope and instrument optics.

### 10.3 Coordination Among Teams

On a large flight project, there are too many details for everyone to be up to date on everything. The main way for different teams to coordinate is through a common set of requirements developed in Phase A of the project. Problems arise when there is a missing requirement on some key aspect. With hundreds of requirements, there is also a danger of some low-level requirements being far-reaching but unknown to everyone who might need to know them.

1. Define requirements on the layouts of the masks and fiducial markings. As mentioned in Sec. 4.1, one of the major issues we encountered was a lack of requirements for the mask layouts. This only became apparent when the final mask design review was approaching and the fabrication team noticed that the layouts were not on the list of deliverables. In the top-down derivation of the project requirements, the focus had been solely on specifying how one mask at each plane would provide the desired contrast, throughput, and other performance metrics. Examples of missing requirements were how many of each mask or fiducial there should be and where.

2. Stricter contrast requirements on HWO will require the observatory and coronagraph instrument to be designed as a whole. Coordination at the mission level is also critical when designing a coronagraph. The ideal situation would be to co-design the observatory and the instrument together because of how sensitive coronagraphs are to the telescope aperture and wavefront stability. The Roman Coronagraph Instrument, as a Class D technology demonstration, is not allowed to place requirements on the observatory. That has mostly been workable for our instrument, however, because of our lower contrast goals, regular communications with the observatory team, and their willingness to take our concerns into account. For the harder task of characterizing earth-like exoplanets with HWO, a balance will need to be found that does not place too much of the performance burden on either the observatory or the coronagraph instrument.

### 10.4 Mask Substrates and Fabrication

1. Use off-the-shelf substrates for masks when possible. One of the lessons learned with the mask substrates is that we should have tried harder to utilize standard-sized optics. This is mostly related to the four 23-mm-diameter fused silica substrates on FPAM. If we had been able to use standard 25.0-mm or 25.4-mm blanks, they could have been off-the-shelf parts. Instead, they were costly with long lead times.

2. Finalize custom substrate sizes before ordering them. The SPAM mask substrate had to be custom fabricated because it was much thicker (4 mm) than a standard silicon wafer for the electronics industry. Unfortunately, we learned that we should have finalized the dimensions before requesting the bare, polished substrates. Most of our SPM fabrication time went to dicing, beveling, and cleaning the mask array to smaller lateral dimensions instead of the planned part of etching the mask designs. This was time-consuming and inefficient compared to having the substrate provided at the correct size to begin with.

3. For masks that are hard to fabricate, try to put them on separate substrates to increase yield. Although it reduced the lateral stage travel needed, making several SPMs per substrate greatly reduced the fabrication yield. The large number of fabrication steps for the SPMs makes the odds of getting four nearly perfect devices very low. For HWO there will need to be a trade study to determine whether it will be more cost-effective to make lone SPMs on smaller substrates or multiple SPMs on just one or two substrates.

4. Build protective guards and/or sleeves into mask mounting assemblies to prevent damage to masks. Another lesson related to the SPMs is that they are fragile and cannot tolerate any contact with the optical surface. For HWO this will be true regardless of which of the two leading technologies is used—either black silicon or carbon nanotubes. The upper left and right black corners of the instrument's SPM array—fortunately outside the SPM beam footprint—were damaged from scuffs during instrument assembly and alignment. In the future, it would be wise to consider mounts with a built-in, protective guard or sleeve near the the substrate to prevent accidents.

### 10.5 Testbed Versus Flight Differences

In mask design as well as other areas such as alignment, calibration, and wavefront correction, problems tended to arise when the flight instrument implementation differed from our past testbed experience on the High Contrast Imaging Testbed (HCIT) Facility at JPL.

1. Determine early on (e.g., in Phase A) what fiducial markings are needed at each mask plane for automated mask alignment and calibration of clocking and magnification. In the HCIT, much of the coronagraphic alignment and calibration before Roman had relied on manual operations by a person in the loop. For Roman, we needed automated routines to guarantee meeting the tight requirements on accuracy and precision. Those algorithms did not exist yet, so we had to include our best guess of which fiducials would be helpful. For HWO, there should be requirements on including fiducials to guarantee that automated, accurate alignment and calibration will be possible.

2. The mask designs may need to be designed for a larger spectral bandwidth because of the finite-width transmission roll-off at bandpass edges. Defining the width of a bandpass seems like a simple problem, but complications arose because we had not been using filters in the HCIT. The HCIT testbeds normally use an NKT Photonics supercontinuum source combined with a SuperK VARIA tunable filter to select the bandpass. For the instrument, we use discrete color filters, which have a roll-off at the bandpass edges with a noticeable difference ($\Delta\lambda/\lambda$ up to 2%) between the bandwidth as defined by the FWHM versus the full width at 90% maximum. To meet the much stricter contrast requirements of HWO, the roll-off of filter transmissions will need to be accounted for early on in the mask designs and color filter choices.

3. Flight instruments must be much cleaner than coronagraphic testbeds because of particulate redistribution during launch. One of the most important differences between the testbeds and the flight instrument is the distribution of particulate contamination. A testbed can be much dirtier because gravity makes dust settle onto the bench below the upright masks. For flight, however, launch vibrations redistribute the particulates evenly within the instrument, so the same cleanliness rating should result in worse performance in orbit than in a testbed. Because the Roman Coronagraph Instrument cannot achieve as high contrast as HWO needs, it can tolerate more, larger particles. Our instrument has been measured to be better than particulate contamination level (PCL) 300. We successfully achieved contrast levels near $10^{-8}$ contrast and did not see bright incoherent spots in the dark hole, so contamination is not likely a limiting factor. For HWO to achieve two orders of magnitude better contrast, more in-depth studies should be performed to follow up on past studies and determine exactly how clean the instrument must be. Initial estimates related to the Terrestrial Planet Finder mission suggest that PCL100 or even PCL50 might be required for the most sensitive optics.[44] In another study directly modeling dust particles on an FPM in a $\approx F/30$ beam, particles as small as

2.5 microns wide can cause incoherent spots in the dark hole as bright as $10^{-8}$ contrast.[45]

4. The only viable way of guaranteeing a focal plane mask without contamination in the most sensitive areas is by requiring a more stringent cleanliness rating. As mentioned in Sec. 6.2, simply including more copies of an FPM is not a viable solution for particulate contamination. If the instrument is not clean enough, it could statistically take tens or hundreds of masks to have one without significantly sized (e.g., 5 to 10 microns across) dust in a key part of the focal plane. However, the only way to tell if there is unresolvably small dust on an FPM in orbit is by performing HOWFSC to high contrast with each FPM until a clean one is found. Realistically, there is only enough time to do this for a handful of masks, so enforcing the cleanliness of the FPM in the first place is the only viable solution.

## 11 Summary

In this paper, we have shown and described all the coronagraphic masks that are included in the Roman Coronagraph Instrument. To assist potential users of the instrument, we have provided diagrams and tables of all the designed mask configurations. As part of the instrument's technology demonstration, there is one required HLC mask configuration for imaging and four best-effort SPC mask configurations for imaging, spectroscopy, and polarimetry. The NASA Exoplanet Exploration Program contributed many unsupported mask configurations, one or more of which might be commissioned if prerequisite maturation is completed and if project resources allow. To aid the design of a coronagraph instrument for HWO, we have detailed the numerous mask design and mask layout choices made for the Roman Coronagraph Instrument. Broad aspects of the layouts such as how many substrates to use and how far the PAMs could move were established early on in the project. That still left flexibility in the mask layouts, especially for the smaller masks in the focal planes. After prioritizing space for the required and best-effort masks, we included the community's highest priority contributed masks to bolster the potential for future expansion of the scientific and technological capabilities of the instrument.

The flight masks were fabricated from 2020 to 2022 in JPL's Microdevices Laboratory using custom substrates provided by JAXA and commercially available SOI wafers. The best of each device was selected for flight, and all were made within the required tolerances—often much better. Two of the mask configurations, the band 1 narrow FOV HLC and wide FOV SPC, received end-to-end testing during the instrument's TVAC testing in the spring of 2024 and successfully reached close to $10^{-8}$ contrast in the allotted time. Pre-TVAC characterizations and cursory TVAC testing of the best-effort spectroscopy SPC indicate that the masks and prisms are in good working order and survived all the preparatory thermal and vibe tests. The next time the required and best-effort masks will be used is during the instrument's commissioning phase, followed by a 90-day observation phase spread over 18 months.

## 12 Appendix A: Color Filters

All of the color filters included in the Roman coronagraph are listed in Table 5. The four broadband filters, to be used after the dark hole is dug, are named 1F, 2F, 3F, and 4F. (The *F* stands for "full" and is generally omitted for brevity.) All the other filters are subsets of the four larger bandpasses and are primarily included for high-order wavefront estimation. There are three subbands each for bands 1 and 4 and five each for bands 2 and 3. Because bands 2 and 3 overlap, though, subbands 3A and 3B are also the two longest wavelength subbands in band 2. The narrowest filters, 2C and 3D, are used for spectroscopy wavelength calibration and line spread function measurements, rather than wavefront correction. Note that filter 2C coincides with the H$\alpha$ spectral line.

Table 5 All color filters and their designed wavelength coverage. The as-built filter properties may differ slightly from these and will be maintained in the list of instrument parameters at Ref. 46, whenever new information becomes available.

| Filter name | Center wavelength (nm) | FWHM (nm) |
| --- | --- | --- |
| 1F | 575 | 58 |
| 1A | 555.8 | 19.2 |
| 1B | 575 | 19.2 |
| 1C | 594.2 | 19.2 |
| 2F | 660 | 112 |
| 2A | 615 | 22 |
| 2B | 638 | 18 |
| 2C | 656.3 | 6.6 |
| 3F | 730 | 122 |
| 3A | 681 | 24 |
| 3B | 704 | 24 |
| 3C | 727 | 20 |
| 3D | 754 | 7.5 |
| 3E | 777.5 | 27 |
| 3G | 752 | 25 |
| 4F | 825 | 94 |
| 4A | 792 | 28 |
| 4B | 825 | 30 |
| 4C | 857 | 30 |
| CLEAR | N/A | N/A |
| DARK | N/A | N/A |

## 13 Appendix B: List of Abbreviations

All acronyms and abbreviations are defined in the main text, but they are defined again below for the reader's convenience.

- ACWG: AFTA Coronagraph Working Group
- AFM: atomic force microscope
- AFTA: Astrophysics-Focused Telescope Asset
- AOI: angle of incidence
- AR: anti-reflective
- Astro2020: Decadal Survey on Astronomy and Astrophysics 2020
- AYO: Altruistic Yield Optimizer
- CFAM: color filter alignment mechanism
- DM: deformable mirror
- DPAM: dispersion polarization alignment mechanism
- EXCAM: exoplanetary systems camera
- ExEP: Exoplanet Exploration Program

- EXOSIMS: Exoplanet Open-Source Imaging Mission Simulator
- FALCO: Fast Linearized Coronagraph Optimizer software package
- FCM: focus control mirror
- FM: fold mirror
- FOV: field of view
- FPAM: focal plane (mask) alignment mechanism
- FPM: focal plane mask
- FSAM: field stop alignment mechanism
- FSM: fast steering mirror
- FWHM: full width at half maximum
- HCIT: high-contrast imaging testbed facility
- HLC12: the substrate on FPAM intended for HLCs in bands 1 and 2
- HLC34: the substrate on FPAM intended for HLCs in bands 2, 3, and 4
- HLC: hybrid Lyot coronagraph
- HOLE: the through-hole position on FPAM
- HOWFSC: high-order wavefront sensing and control
- HWO: Habitable Worlds Observatory
- IWA: inner working angle
- JAXA: Japan Aerospace Exploration Agency
- JPL: Jet Propulsion Laboratory
- LOBE: LOWFS Optical Barrel Element
- LOCAM: low-order (wavefront sensing) camera
- LOS: line of sight
- LOWFSC: low-order wavefront sensing and control
- LSAM: Lyot stop alignment mechanism
- MDL: microdevice laboratory
- MPIA: Max Planck Institute for Astronomy
- MSWC: multi-star wavefront control; also the SPAM position for that purpose
- NASA: National Aeronautics and Space Administration
- ND: neutral density
- ND225: the FPAM position for the neutral density filter with optical density 2.25
- ND475: the FPAM and FSAM position for the neutral density filter with optical density 4.75
- NFOV: the SPAM and LSAM position for narrow field of view (that is, HLCs)
- OAP: off-axis parabola
- OIP: optical interface plate
- OWA: outer working angle
- PAM: precision alignment mechanism
- PCL: particulate contamination level
- PIAACMC: phase-induced amplitude apodization complex mask coronagraph
- PMGI: polydimethylglutarimide
- Pointing, Positioning, Phasing and Coordinate Systems (PPPCS)
- PROX-E: proximity electronics
- PSF: point spread function
- Roman: Nancy Grace Roman Space Telescope
- RV: radial velocity
- SNR: signal-to-noise ratio
- SOI: silicon on insulator

- SPAM: shaped pupil alignment mechanism
- SPEC: the SPAM and LSAM position for contributed, nominal-orientation spectroscopy
- SPECROT: the SPAM and LSAM position for unsupported, rotated spectroscopy
- SPC: shaped pupil coronagraph
- SPC12: the substrate on FPAM intended for SPCs in bands 1 and 2
- SPC34: the substrate on FPAM intended for SPCs in bands 3 and 4
- SPM: shaped pupil mask
- TRL: technology readiness level
- TTR5: technology threshold requirement 5
- TVAC: thermal vacuum
- WFOV: the SPAM and LSAM position for wide field-of-view SPCs
- ZWFS: Zernike wavefront sensor

## Disclosures

The authors declare that there are no financial interests, commercial affiliations, or other potential conflicts of interest that could have influenced the objectivity of this research or the writing of this paper.

## Code and Data Availability

The two-dimensional, numerical representations of the flight masks are available as part of the instrument simulator software package CGISim on SourceForge at https://sourceforge.net/projects/cgisim/. The release of code used in this work is governed by JPL and NASA rules which limit dissemination scope and methodology. The corresponding author should be contacted about the availability of any code or data used in this paper.

## Acknowledgements

This work was performed in part at the Jet Propulsion Laboratory, California Institute of Technology, under contract with the National Aeronautics and Space Administration (NASA). This article is an expanded update of the SPIE conference proceeding "Flight mask designs of the Roman Space Telescope Coronagraph Instrument.[47]"

**A. J. Eldorado Riggs** is an optical engineer at NASA's Jet Propulsion Laboratory. His research focuses on the high-contrast imaging of exoplanets, in particular, mask optimization and wavefront sensing and control for the Coronagraph Instrument on the Nancy Grace Roman Space Telescope. He received his BS degree in physics and mechanical engineering from Yale University in 2011 and his PhD in mechanical and aerospace engineering from Princeton University in 2016.

**Kunjithapatham Balasubramanian** has been a senior optical engineer at the Jet Propulsion Laboratory, California Institute of Technology, since 2004. With a PhD in optical sciences from the University of Arizona in 1988, he conducted research on optical materials, thin films, and micro/nanostructured optical devices. He led the development of exoplanet coronagraph masks for testbeds at JPL, Princeton University, NASA GSFC, and ARC. He now supports further research and development of advanced optical devices and coronagraph masks with the JPL team.

**Tyler D. Groff** received his bachelor's degree in mechanical engineering and astrophysics in 2007 from Tufts University and his PhD in mechanical and aerospace engineering from Princeton University in 2012. His research focuses on visible and near-infrared instrumentation for exoplanet science, coronagraph design, and wavefront control. He is the lead engineer at Goddard for the Roman Coronagraph Instrument prism and polarizer modes. The optical verification lead for the Roman wide field instrument, and the principal investigator for the parabolic deformable mirror and CHARIS IFS at Subaru telescope.

**Brian Monacelli** is a principal optical engineer and technical group supervisor at the Jet Propulsion Laboratory, where he led the optical assembly and alignment of the Coronagraph Instrument for the Roman Space Telescope and the SHERLOC Instrument of the Mars2020 Perseverance Rover, among other optical instruments. He received his BS degree in applied optics from Rose-Hulman Institute of Technology, his MS degree in optics from the University of Rochester, and his PhD in optics from the University of Central Florida/CREOL. He writes applied curricula that teach technicians to learn hands-on laboratory skills via the Pasadena City College Laser Technology program. He also serves on the Optics and Electro-Optics Standards Council, where he supports the development of optics standards.

**Erkin Sidick** received a BS degree in electrical engineering from Xinjiang University, Urumchi, Uyghur Autonomous Region, China, in 1983, an MS degree in physics from California State University, Northridge, in 1990, and a PhD in electrical engineering from the University of California, Davis, in 1995. After conducting research jointly at UC Davis and Sandia National Lab for 1.5 years, and working 7 years in three different optics companies in Silicon Valley, California, he joined JPL in 2004 as a senior optical engineer. There he worked on a variety of projects related to space telescopes, conducting optical modeling and simulations in the areas of wavefront sensing and control as well as integrated modeling. He recently started at Reflect Orbital in California as an optical modeling engineer.

**Nicholas Siegler** is an astrophysicist and the chief technologist for NASA's Exoplanet Exploration Program located at the Jet Propulsion Laboratory in Pasadena, CA. With collaborators from all over the country, he helps identify and mature technologies needed to enable possible future NASA missions to look for evidence of life on planets outside of our solar system. He helped found the NASA Starshade Technology Development Activity and currently manages the high-contrast imaging testbed facility testing next-generation coronagraphs. Nick also has experience in systems engineering and project management on numerous projects.

**Neil T. Zimmerman** is a research astrophysicist in the Exoplanets and Stellar Astrophysics Laboratory at NASA's Goddard Space Flight Center. He is a member of the Nancy Grace Roman Space Telescope Project Science team and works on simulations and tests of instrument technologies for exoplanet imaging and spectroscopy.

Biographies of the other authors are not available.